\documentclass[reprint,amsmath,amssymb,aps,twocolumn,prx]{revtex4-2}

\usepackage{graphicx}% Include figure files
\usepackage{bm, slashed}% bold math
\usepackage{tikz} % tikz package
\usepackage{hyperref}% add hypertext capabilities
\usepackage{comment}
\usepackage{mathdots,bbold}
\usepackage{soul}
\usepackage{braket}
\usepackage[normalem]{ulem}
\setcitestyle{square,numbers}
\usepackage{pgfplotstable}
\usepackage{zref-savepos}  % Track positions to detect left/right column
\usepackage{etoolbox}      % Needed for patching
\usepackage[left]{lineno}

\makeatletter
\newcommand{\fixcolumnlinenos}{
  \zsavepos{lineno-column-check}
  \ifdim\zposx{lineno-column-check} < 0.5\textwidth
    \renewcommand\makeLineNumber{\llap{\linenumberfont\LineNumber\hspace{0.5em}}} % Left column: left align
  \else
    \renewcommand\makeLineNumber{\hfill\linenumberfont\LineNumber} % Right column: right align
  \fi
}
\preto\linenumbers{\fixcolumnlinenos} % Apply column check before line numbers
\makeatother

\begin{document}
%\linenumbers

%Temptative title
\title{An efficient finite-resource formulation of non-Abelian lattice gauge theories beyond one dimension}

\author{Pierpaolo Fontana$^{1,2}$}
\email{Pierpaolo.Fontana@uab.cat}
\author{Marc Miranda-Riaza$^{1}$}
\author{Alessio Celi$^{1}$}

\affiliation{$^{1}$Departament de Física, Universitat Autònoma de Barcelona, 08193 Bellaterra, Spain}
\affiliation{$^{2}$ICFO - Institut de Ciències Fotòniques, The Barcelona Institute of Science and Technology, 08860 Castelldefels (Barcelona), Spain}

\date{\today}

\begin{abstract}
Non-Abelian gauge theories provide the most accurate description of fundamental interactions, showing remarkable agreement with experimental data in cosmology and particle physics. Highly precise predictions can be made using standard techniques, both in the continuum and in the lattice frameworks. However, classical methods have limitations, particularly when attempting to extrapolate the continuum limit from the study of lattice gauge theories. Complementing classical computations or combining them with quantum computational methods, to improve the predictions towards the continuum limit with current quantum resources, is a formidable open challenge. In this paper, we propose a resource-efficient method to compute the running of the coupling in non-Abelian gauge theories beyond one spatial dimension. We first represent the Hamiltonian on periodic lattices in terms of loop variables and conjugate loop electric fields, exploiting the Gauss law to retain the gauge-independent ones. Then, we identify a local basis for small and large loops {\it variationally} to minimize the truncation error while computing the running of the coupling on small tori. Our method enables computations at arbitrary values of the bare coupling and lattice spacing with current quantum computers, simulators and tensor-network calculations, in regimes otherwise inaccessible.
\end{abstract}

\maketitle

\section{\label{intro}Introduction}
Recent progress in analog and digital quantum simulators \cite{feynmanQS1982,Cirac-Zoller,Dalibard_UCatoms2012,Blatt_trappedions2012, rmpNori} of condensed matter models \cite{maciejUltracold} has spurred renewed interest \cite{ZoharPRL2012,Banerjee2012,tagliacozzo_NC2013,zohar2015quantum,wiese,Dalmonte-Montangero,Preskill2018Nov,barros2020gauge,banulsRPP2020,Banuls2020,aidelsburger2022,davoudi_QSforHEP2022,Halimeh2023Oct,DiMeglio2023Jul} in Hamiltonian lattice gauge theory (LGT) \cite{Kogut-Susskind}. These simulators can reproduce the real-time evolution of a target model and handle fermions, either by exploiting the fermionic constituents of ancillary quantum systems or by means of a polynomial number of quantum gates. This capability allows them to avoid the infamous “sign problem” \cite{troyer_wiese2005,gattringer2009,fukushima2010,Delia2019}, which hinders real-time and high-fermion-density Monte Carlo computations for Lagrangian LGTs in Euclidean space. In one spatial dimension (1D), experimental implementations of valuable toy models of low-dimensional LGTs with different quantum platforms have achieved remarkable success \cite{Martinez2016,KlcoPRA2018,Schweizer2019,Yang2019,Kokail2019,Mil2020,SuracePRX2020,KlcoPRD2020,NguyenPRXQuantum2022,ZhaoScience2022,Mildenberger2022,SuPRR2023,Zhang2023preprint}. Proposals for two-dimensional (2D) systems have also shown promise, encompassing the simulation of emerging gauge theories in condensed matter systems \cite{Wen,Celi2019,Yang2019,SuracePRXQuantum,Sachdev_2023} and $U(1)$ gauge theories relevant to particle physics \cite{Haase2021resourceefficient, PaulsonPRX2021, BauerPRD2023}. A first proof of principle experiment has been very recently realized with qudits \cite{Meth2025}, showcasing the growing potential of quantum simulations and computations. However, we are still far from performing full-scale quantum computations, such as scattering processes on quantum processors \cite{Jordan2012,Jordan2014}, in real-world gauge theories, like quantum chromodynamics (QCD). This challenge is further complicated by the matter content of the Standard Model \cite{Peskin,Scwhartz} and the huge energy scales involved in particle collisions \cite{BauerPRL2021}. Even before including these aspects, pure non-Abelian LGTs in 2D and 3D already exceed the capabilities of current quantum hardware \cite{LammPRD2019,NachmanPRL2021,LammPRL2022}. To bridge this gap, we propose here a resource-efficient approach to simulate non-Abelian LGTs in higher dimensions.
%%%%%%%%%%%%%%%%%%%%%%%%%%%%%%%%%%%%%
\begin{figure*}[t!]
    \centering
    \includegraphics[width=0.95\linewidth]{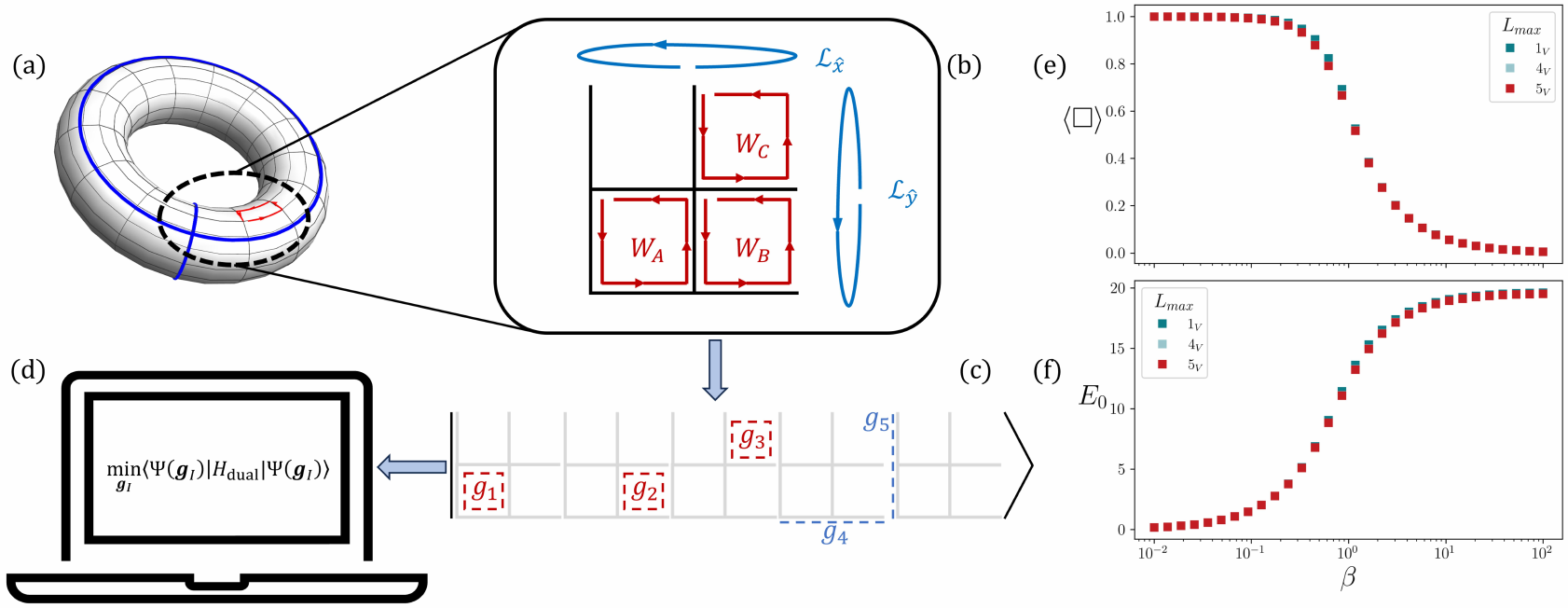}
    \caption{(a) Arbitrary periodic lattice represented as the surface of a torus in the three-dimensional space. The fundamental degrees of freedom after the change of variables to the loop (or dual) degrees of freedom are the plaquettes (red square) and the two large loops wrapping the torus (blue circles). (b) Unfolded minimal torus, i.e., the single periodic plaquette, identified by four sites ($\bullet$) and eight links ($-$). In the dual formulation, the physical states of the pure gauge theory are represented by three independent plaquettes (red loops) and two large loops (blue loops). (c) Cartoon picture of the basis for the physical states of the lattice gauge theory on the minimal torus. Every fundamental degree of freedom is associated to a free parameter $g_I$, $I=1,\ldots,5$, that we determine \textit{variationally} (d) by minimizing the expectation value of the dual Hamiltonian in the local parameters space. (e) Expectation values of the plaquette operator $ \langle\square\rangle$ can be computed for all values of the coupling constant $\beta=(2g^2)^{-1}$ with a small number of states. Fast convergence is also appreciated in the energy of the ground state (f).}
    \label{TeaserVariational}
\end{figure*}
%%%%%%%%%%%%%%%%%%%%%%%%%%%%%%%%%%%%%

For continuous gauge groups, the local Hilbert space associated with a single gauge degree of freedom, which is a link in the usual Wilson formulation \cite{Wilson1974}, is infinite dimensional because the electric field is unbounded. To engineer a feasible quantum simulator, the link degrees of freedom must be truncated to a finite Hilbert space in a chosen basis. The most common choice is the \textit{electric basis}, where the basis states are eigenvalues of the electric field squares. Here the electric interactions and the Gauss laws, i.e., the local conservation laws defining the gauge symmetry, are diagonal. The truncation of the local Hilbert space in this basis led to quantum link models \cite{HORN1981,Orland1990,Chandrasekharan1997}, finite spin models interesting {\it per se} as emerging gauge theories in condensed matter, like quantum spin ice \cite{quantumSpinIceReview}. However, in particle physics, one must take the continuum limit, which requires considering, beyond 1D, the regime where the magnetic interactions dominate, that is, the weak coupling regime. Since magnetic interactions are completely off-diagonal in the electric basis, this truncation scheme becomes increasingly costly -- to keep the same precision at smaller couplings, one must retain more and more states, making it impractical for current quantum hardware or for tensor-network algorithms. In this regime, it would be more convenient to use the basis where the link operators are diagonal, ensuring that the magnetic part of the Hamiltonian is diagonal as well. However, in this basis the Gauss law is not diagonal, which means that truncations inherently break gauge symmetry.

In this work we reformulate the Hamiltonian of non-Abelian gauge theories directly in terms of a specific set of gauge invariant loops degrees of freedom. We separate the dual Hamiltonian into a local part -- equivalent to the sum of single plaquette Hamiltonians -- and non-local terms \cite{Ligterink2000}. In order to deal with the infinite-dimensional local Hilbert space, we propose a truncation scheme based on the eigenbasis of the local Hamiltonian. We boost this approach by implementing a physically motivated variational procedure to select the optimal truncation basis. We benchmark our approach by determining the running of the coupling — the expectation value of the plaquette operator $ \langle\square\rangle$ at different couplings \cite{CreutzSU(2)1980} — of $SU(2)$ LGT on a minimal torus in 2D. We demonstrate a percent level precision in the determination of the ground state energy and $ \langle\square\rangle$ by retaining a number of states available in current quantum computation and simulation experiments \cite{Meth2025}. Our approach is summarized graphically in Fig. \ref{TeaserVariational}. 

We first reformulate the Kogut--Susskind (KS) $SU(N)$ Hamiltonian on a torus in terms of loop variables and conjugate loop electric fields, by means of canonical transformations on the initial links and electric fields -- a procedure already introduced in \cite{MathurPRD2015} and applied on lattices with open boundary conditions. Through the explicit resolution of all Gauss laws, we write the dual Hamiltonian for physical states only on lattices with periodic boundary conditions (PBC), to minimize boundary effects, and no constraints left over. We write explicitly the reformulation for the minimal torus and generalize it for arbitrary lattice sizes. The dualization introduces non-localities in the electric part, while it simplifies the magnetic part of the Hamiltonian. The Hilbert space of the selected loop variables, i.e., the dual degrees of freedom, naturally fits the basis where magnetic interactions are diagonal. We highlight that this is not equivalent to the dualization of the theory in the space of \textit{all possible} loops: the gauge fixing procedure prevents the problem of the over-completeness of the loop basis, avoiding the presence of the Mandelstam constraints \cite{MandelstamPR68}. We exemplify the basis of the local Hamiltonian using a specific set of coordinates in the gauge group, and then construct a complete basis by taking the tensor product of local bases \cite{Ligterink2000}. This procedure enables an optimized representation that adapts to different coupling regimes. While previous studies have explored the digitization of the link degree of freedom through the discretization of the group manifold \cite{Bauer2023}, the use of a variationally optimized local eigenbasis allows us to reduce significantly computational costs, while retaining high accuracy. Moreover, since the corresponding matrix elements can be efficiently computed using classical methods, the overall resource requirements for both classical and quantum computations are significantly reduced. The computational efficiency of this basis construction, in terms of retained resources, is exemplified in Table \ref{table1}.

\begin{table*}[t!]
    \label{table1}
    \setlength{\tabcolsep}{12pt}
    \begin{tabular}{c||ccc}
        $\beta$ & \begin{tabular}[c]{@{}c@{}}Standard truncation\\  (electric basis)\end{tabular} & 
        \begin{tabular}[c]{@{}c@{}}Reformulation\\ (interpolating basis)\end{tabular} & \begin{tabular}[c]{@{}c@{}}Variational\\ (interpolating basis)\end{tabular} \\ \hline\hline
        0.01 & 1                                                                               & 1                                                                             & 1                                                                           \\
        1    & 2744                                                                            & 64                                                                            & 64                                                                          \\
        100  & $>$2744                                                                           & 125                                                                          & 1                                                                          
    \end{tabular}
    \caption{Computational cost for different approaches. We estimate the number of states required to reach a 1\% accuracy in the ground state energy of the minimal torus for the pure $SU(2)$ LGT (see Sec. \ref{work_basis_periodicplaqPBC}) for different basis and truncations. The three columns refer to the standard electric representation (leftmost column), our approach using the single plaquette basis, described in Sec. \ref{CT_plaquette_PBC} (central column), and our optimized local basis (rightmost column), in which the values of the local couplings are variationally determined through the strategy described in Sec. \ref{variational_GS}.}
\end{table*}

While there are alternative approaches to the electric basis that require finite resources (for a review of available reformulation in 1D see \cite{DavoudiPRD2021}), they are different from the one presented here and have not been used to compute the running of the coupling in non-Abelian gauge theories. There are proposals to approximate the KS Hamiltonian for continuous gauge groups with the one for discrete groups \cite{KuhnPRA2014,ZoharPRD2015,notarnicola_JPA2015,ErcolessiPRD2018}, and to combine it with improved Hamiltonians \cite{LammPRL2022} to enhance the weak coupling behavior, or with deformations of the algebra through quantum groups \cite{ZachePRL2023}. Alternatively, employing the electric basis, Ref. \cite{brower_wiese_PRD1999} proposes to achieve the continuum limit by using quantum link models with one extra dimension and performing dimensional renormalization. Several works push forward the dual formulation of Abelian \cite{Haase2021resourceefficient,Bender-Zohar,Fontana2022:LGT,BauerPRD2023} and non-Abelian gauge theories for discrete \cite{DanielGCPRL2022,DiscreteLammPRD2022,Zache2023fermionquditquantum,PradhanPRD2023,BalliniPRD2024} and continuous \cite{StrykerPRR2020,StrykerPRD2020,MendicelliPRD2021,Cataldi2023,StrykerPRD2023,HalimehGrusdt2024,Ciavarella2024Feb,Calajo2024Feb} gauge groups to simplify the form of the Hamiltonian and achieve an exponential reduction of quantum resources needed for simulation. Refs. \cite{Haase2021resourceefficient,BauerPRD2023} employ this formulation to compute the running of the coupling in 2D quantum electrodynamics, while \cite{Garofalo2023Nov} proposes the simulation of $SU(2)$ Schwinger-like model in $d=1$.

% Organization of the paper
The paper is organized as follows. In Sec. \ref{KS_Hamiltonian}, we remind the properties of $SU(N)$ lattice gauge theories and define the fundamental variables and the KS Hamiltonian. In Sec. \ref{CT_plaquette_PBC}, we present our dualization procedure and the encoded Hamiltonian and detail them for the minimal torus (for details on generic tori see Appendix \ref{generalization_square_lattice}). In Sec. \ref{H_singleperiodicplaq_PBC}, we discuss the general features of the dual Hamiltonian dynamics. We then consider the specific case of $SU(2)$ in Sec. \ref{work_basis_periodicplaqPBC}, where we show how to optimally choose the local basis through a variational procedure, and compute the expectation value of the running of the coupling in the variational ground state. We discuss our results and the perspectives they open  in Sec. \ref{conclusions}. In the Appendices we present all the numerical and computational details, and the generalization to larger tori of arbitrary sizes.

%%%%%%%%%%%%%%%%%%%%%%%%%%%%%%%%%%%%%%%%%%%%%%%%%%%%%%%%%%%%%%%%%%%%%%%%%%%%

\section{\label{KS_Hamiltonian}$SU(N)$ lattice gauge theories}
Throughout this paper, we consider the Hamiltonian formulation of a pure $SU(N)$ non-Abelian LGT, on a square lattice $\Sigma$ of size $N_x\times N_y$ with PBC. Each site of the lattice is denoted by $\bm{n}=(n_x,n_y)$, where $n_\mu\in\{0,\ldots,N_{\mu}-1\}$ and $\mu\in\{\hat{x},\hat{y}\}$. In the KS formulation of LGTs \cite{Kogut-Susskind}, the gauge fields reside on the links of the lattice and provide the parallel transport of the color charge. The link variables are denoted by $U_{\mu}(\bm{n})\in SU(N)$, as shown in Fig. \ref{basic_CT + plaquette_changeofvars}(a), where the site $\bm{n}$ and direction $\mu$ are used to identify them within the lattice. Under local transformations $\Omega (\bm{n})\in SU(N)$, the gauge field  transforms as
\begin{equation}
    U_{\mu}(\bm{n}) \rightarrow \Omega(\bm{n}+\mu) U_{\mu}(\bm{n})\Omega^{\dagger}(\bm{n}).
    \label{wilson_line_transform}
\end{equation}
The gauge transformation in Eq. \eqref{wilson_line_transform} is realized by two independent $\mathfrak{su}(N)$ algebras associated to each link. These are the left and right electric fields, $E^a_{L(R),\mu}$, which are conjugate to the gauge field and satisfy the commutation relations 
\begin{equation}
     [E^a_{L,\mu}(\bm{n}),U_{\mu}(\bm{n})] = -T^aU_{\mu}(\bm{n}),
     \label{left_CR}
\end{equation}
\begin{equation}
     [E^a_{R,\mu}(\bm{n}+\hat{\mu}),U_{\mu}(\bm{n})] = U_{\mu}(\bm{n})T^a,
     \label{right_CR}
\end{equation}
where the group generators $T^a$ satisfy the defining commutation relations $[T^a,T^b]=if^{abc}T^c$ (see Appendix \ref{appendix_YM_action}). By introducing the contraction of the electric fields with the group generators, i.e.,  $E_{L(R),\mu}(\bm{n})=E^a_{L(R),\mu}(\bm{n})T^a$, where repeated indices are summed over, the left and right electric fields are related through
\begin{equation}
    E_{R,\mu}(\bm{n}+\hat{\mu})=-U^\dagger_{\mu}(\bm{n})E_{L,\mu}(\bm{n})U_{\mu}(\bm{n}),
    \label{single_parallel_transport}
\end{equation} and thus satisfy $\text{Tr}\;E^2_{\mu}(\bm{n})=\text{Tr}\;E^2_{L,\mu}(\bm{n})=\text{Tr}\;E^2_{R,\mu}(\bm{n})$, i.e., they have equal quadratic Casimir operators. In components, Eq. \eqref{single_parallel_transport} can be written by introducing the orthogonal matrix $\mathcal{R}^{ab}\in SO(N^2-1)$ such that
\begin{gather}
    \nonumber
    E^a_{R,\mu}(\bm{n}+\hat{\mu})=-\mathcal{R}^{ab}(U)E^b_{L,\mu}(\bm{n}),\\ 
    \mathcal{R}^{ab}(U)\equiv2\text{Tr}[U^\dagger T^a U T^b],
    \label{Rab_definition}
\end{gather}
where repeated indices are implicitly summed over. This is nothing but the link in the adjoint representation of $SU(N)$, satisfying the orthogonality condition $\mathcal{R}^{ac}\mathcal{R}^{bc}=\delta^{ab}\mathbb{1}$.

The Gauss laws, for every site $\bm{n}\in\Sigma$, can be written as
\begin{equation}
    \mathcal{G}^a(\bm{n}) = \sum_{\mu=\hat{x},\hat{y}}[E^a_{L}(\bm{n})+E^a_{R}(\bm{n})]=0,
\end{equation}
using the sign convention introduced in Eqs. \eqref{left_CR}, \eqref{right_CR} and \eqref{single_parallel_transport}.

Ultimately, the dynamics of the theory is ruled by the KS Hamiltonian \cite{Kogut-Susskind}, given by
\begin{align}
    \nonumber
    H&=H_E+H_B\\
    &=g^2\sum_{\bm{n},\mu}\text{Tr}\;E^2_{\mu}(\bm{n})+\frac{1}{2g^2}\sum_{P}\text{Tr}[2-(U_P+U_P^\dagger)],
    \label{initial_KS_Hamiltonian}
\end{align}
where $g$ is the coupling constant and $U_P$ are the so-called plaquette operators, defined as the Wilson loops along the individual plaquettes, i.e. $U_P\equiv U_{\hat{x}}(\bm{n})U_{\hat{y}}(\bm{n}+\hat{x})U^{\dagger}_{\hat{x}}(\bm{n}+\hat{y})U^{\dagger}_{\hat{y}}(\bm{n})$. The first term in the Hamiltonian sums the electric energy contribution of each link in the lattice, while the second term incorporates the magnetic energy of all plaquettes.

%%%%%%%%%%%%%%%%%%%%%%%%%%%%%%%%%%%%%%%%%%%%%%%%%%%%%%%%%%%%%%%%%%%%%%%%%%%%

\section{\label{CT_plaquette_PBC}Dualization and encoding for the minimal torus}
\begin{figure}[t!]
    \centering
    \includegraphics[width=1.0\linewidth]{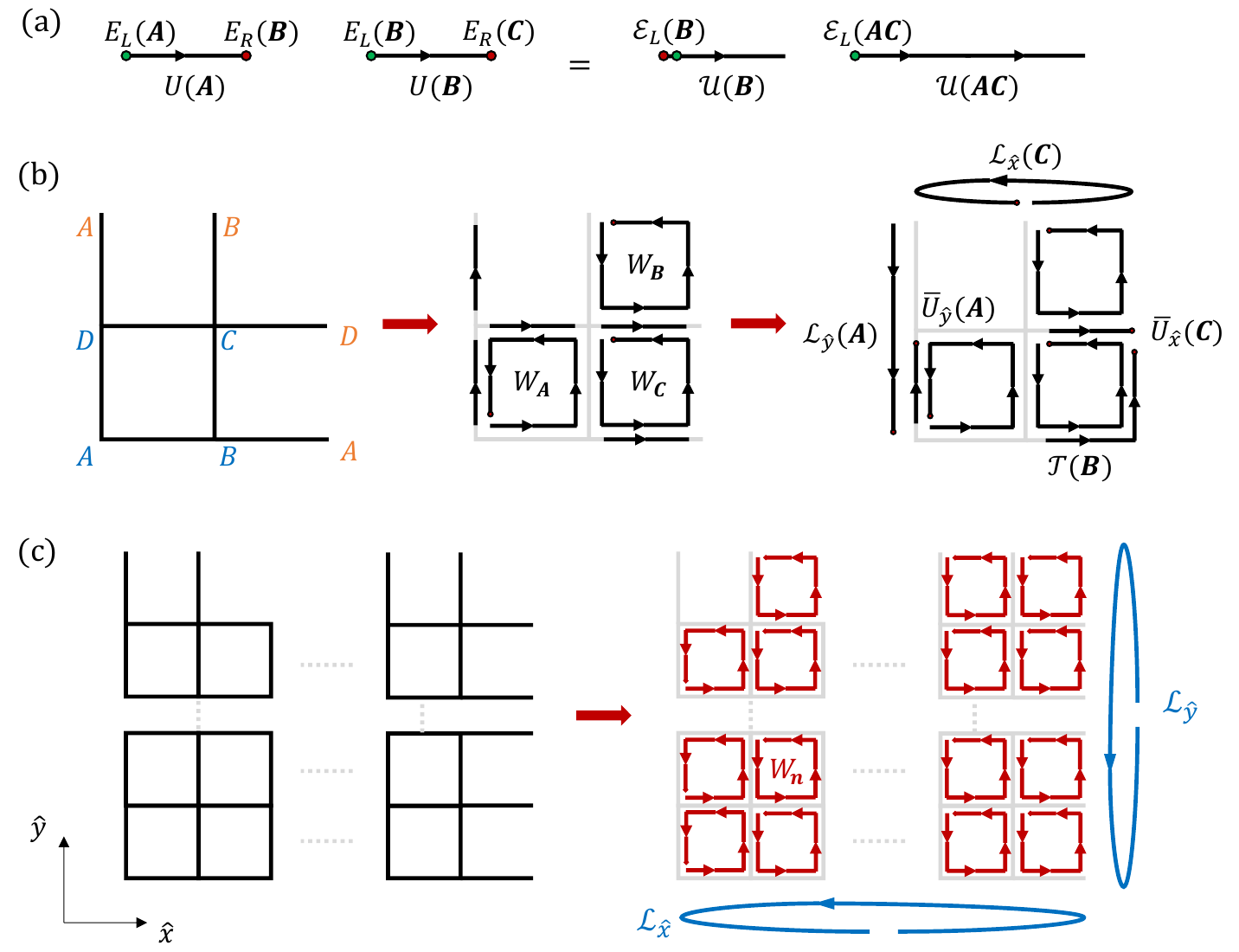}
    \caption{(a) Basic canonical transformation to join two consecutive links, $U(\bm{A})$ and $U(\bm{B})$, connecting $\bm{A}\rightarrow\bm{B}$ and $\bm{B}\rightarrow\bm{C}$, respectively. From $U(\bm{A})$ and $U(\bm{B})$ with conjugate electric fields reported in the left-hand side of the graphical equality, two new links $\mathcal{U}(\bm{B})$ and $\mathcal{U}(\bm{AC})$ are introduced, which are still mutually independent. Here and in the following, we use green (red) dots to represent left (right) electric fields. (b) Graphical representation of all the canonical transformations performed in the main text. The left panel shows the initial Kogut--Susskind links. Blue letters represent physical sites, while orange letters are repeated to explain the lattice periodicity. The central panel shows the results of the first set of canonical transformations, i.e., five links and the three independent plaquettes, and for the sake of simplicity we report only the nomenclature for the plaquettes. The right panel is the final outcome after the second set of canonical transformations, i.e., five closed loops (three plaquettes, two large loops) and three open strings ending in the reference point $\bm{D}$. (c) Extension of canonical transformation to a larger torus. The left panel shows the initial links, while in the right panel we report the independent plaquettes (red loops) and the large loops $\mathcal{L}_{x,y}$ (blue loops).}
    \label{basic_CT + plaquette_changeofvars}
\end{figure}
We now present the reformulation of non-Abelian $SU(N)$ LGTs in terms of physical independent degrees of freedom and detail it for the minimal torus, i.e., the single plaquette with PBC, for which $N_x=N_y=2$. We stress that, as depicted in Fig. \ref{basic_CT + plaquette_changeofvars}(b), the single periodic boundary plaquette contains four physical sites and eight links: four bulk links and four boundary links. The extension to arbitrary tori of size $N_x\times N_y$ is presented in Appendix \ref{generalization_square_lattice}. Our computations generalize the dualization procedures known both in the Abelian case with PBC \cite{Haase2021resourceefficient} and the non-Abelian case with open boundaries \cite{MathurPRD2015,Bauer2023}. As a result of having periodic lattices, we are able to eliminate residual constraints and reduce finite-size effects, which are essential aspects for practical quantum simulations and computations. While these differences may appear technical with respect to the case of open boundary conditions, they require careful consideration as they significantly impact the reformulation of the theory.

As introduced in Sec. \ref{KS_Hamiltonian}, the initial degrees of freedom are the left and right electric fields $E^a_{L,R}(\bm{n})$ and the conjugate $SU(N)$ link variables $U_\mu(\bm{n})$. The main idea behind the reformulation is to realize the change of variables 
\begin{equation}
\begin{split}
    \{E^a_{L,R}(\bm{n}),& U_\mu(\bm{n})\}\rightarrow \\ \{\mathcal{E}^a_{L,R}(\bm{n}),\mathcal{W}(\bm{n}) \}& \oplus\{\mathrm{E}^a_{L,R}(\bm{n}),\mathrm{T}(\bm{n})\},
    \label{changeofvars_general}
\end{split}
\end{equation}
where the first set on the right-hand side contains closed strings (loops) $\mathcal{W}(\mathbf{n})$, and conjugate electric fields $\mathcal{E}^a_{L,R}(\mathbf{n})$, while the second one contains open strings $\mathrm{T}(\mathbf{n})$ and their conjugate electric fields $\mathrm{E}^a_{L,R}(\mathbf{n})$. The $\{\mathcal{E}^a_{L,R}(\bm{n}),\mathcal{W}(\bm{n}) \}$ are the new physical variables, while the set $\{\mathrm{E}^a_{L,R}(\bm{n}),\mathrm{T}(\bm{n})\}$ decouples from the physical one in the dynamics of the theory \cite{Haase2021resourceefficient,Mathur_PLB2015,MathurPRD2015,Mathur_PRD2023}.
In the remainder of the Section we summarize the main steps to realize Eq. \eqref{changeofvars_general}, leading to Eqs. \eqref{inverse_links} and Eqs. \eqref{inverse_E_1}, \eqref{inverse_E_2}, \eqref{inverse_E_3}, \eqref{inverse_E_4} for the inverse relations of links and conjugate electric fields, respectively. We point out, whenever is needed, how we solve the main complications related to the non-Abelian nature of the gauge group. The full dualization procedure, including all the technical details and intermediate steps, is reported in Appendix \ref{appendix_CT_dualization}.

The first step to perform concretely the dualization of Eq. \eqref{changeofvars_general} is the introduction of canonical transformations (CTs) for the links and electric fields \cite{Mathur_PRD2023}. As showed in Fig. \ref{basic_CT + plaquette_changeofvars}(a), starting from two consecutive links $U(\bm{A})$, $U(\bm{B})$ the new independent links are defined as
\begin{equation}
    \mathcal{U}(\bm{B})=U(\bm{B}),\qquad\mathcal{U}(\bm{AC})=U(\bm{A})U(\bm{B})
    \label{basic_CT_links}
\end{equation}
and the conjugate left electric fields are respectively given by
\begin{equation}
    \mathcal{E}_L(\bm{B})=E_R(\bm{B})+E_L(\bm{B}),\qquad \mathcal{E}_L(\bm{AC})=E_L(\bm{A}).
    \label{basis_CT_electricfields}
\end{equation}
The corresponding right electric fields are obtained by parallel transport of the $\mathcal{E}_L$ along the new independent paths, according to the relation 
\begin{equation}
    \mathcal{E}_{R,\mu}(\bm{n}+\mu)=-\mathcal{U}^\dagger_\mu(\bm{n})\mathcal{E}_{L,\mu}(\bm{n})\mathcal{U}_\mu(\bm{n}),
    \label{relation_left_right_electricfields}
\end{equation}
which holds for a general path $\mathcal{U}$ with attached electric fields $\mathcal{E}_{L,R}$, generalizing Eq. \eqref{single_parallel_transport}. We emphasize that such CTs preserve the canonical commutation relations in Eqs. \eqref{left_CR}, \eqref{right_CR} and provide a linear map for the realization of Eq. \eqref{changeofvars_general}.

As a second step, we construct the elementary loops $W_{\bm{n}}$ in Fig. \ref{TeaserVariational}(b) by applying iteratively the CTs to the initial degrees of freedom. By looking at the left panel of Fig. \ref{basic_CT + plaquette_changeofvars}(b), in the original formulation we have eight KS link variables, with three independent Gauss laws, i.e. $\mathcal{G}(\bm{n})=0$ for $\bm{n}=\bm{A},\bm{B},\bm{C}$, since the Gauss law in $\bm{D}$ is not independent due to the topology of the torus  (the total electric flux on a compact manifold is zero). We have then \textit{five independent links}. In the dual formulation, in analogy with the Abelian case \cite{Haase2021resourceefficient}, we end up in three plaquette loops $W_{\bm{A},\bm{B},\bm{C}}$, as due to the PBC the loop $W_{\bm{D}}$ is given by the product of the other three (the magnetic flux through a compact manifold is zero), and two loops wrapping the lattice, that we label $\mathcal{L}_{\hat{x},\hat{y}}$, as depicted in the central and right panels of Fig. \ref{basic_CT + plaquette_changeofvars}(b). Since the fundamental CT in Eq. \eqref{basic_CT_links} preserves the number of links, after the whole procedure we have three open strings left, to be chosen accordingly in our iterative applications of CTs, as we have a certain freedom to recombine the final unpaired links. By taking into account the three Gauss laws for the electric fields conjugate to the plaquette and loop variables, we end up in \textit{five independent loop variables}.

After the whole procedure, we are left with linear invertible relations between the KS variables and the dual ones, essentially realizing Eq. \eqref{changeofvars_general}. We summarize here the final form of the inverse relations for the links
\begin{gather}
    \nonumber
    U_{\hat{x}}(\bm{A})=W_{\bm{A}}\mathcal{L}^\dagger_{\hat{x}}(\bm{C})W_{\bm{C}}, \enspace U_{\hat{y}}(\bm{C})=\mathcal{L}^\dagger_{\hat{y}}(\bm{A})W^\dagger_{\bm{B}},\\
    U_{\hat{y}}(\bm{B})=W^\dagger_{\bm{C}}, \enspace U_{\hat{y}}(\bm{D})=\mathcal{L}^\dagger_{\hat{y}}(\bm{A}),
    \label{inverse_links}\\
    \nonumber
    U_{\hat{x}}(\bm{D})=\mathcal{L}^\dagger_{\hat{x}}(\bm{C}), \enspace U_{\hat{x}}(\bm{C})=U_{\hat{x}}(\bm{B})=U_{\hat{y}}(\bm{A})=\mathbb{1}_{\mathcal{G}}
\end{gather}
and their conjugate electric fields
\begin{gather}  
    \nonumber
    E_{L,\hat{x}}(\bm{A})=\mathcal{E}_L(\bm{A}),\qquad E_{R,\hat{y}}(\bm{B})=\mathcal{E}_L(\bm{B}),\\
    E_{R,\hat{y}}(\bm{C})=\mathcal{E}_L(\bm{C})+\mathcal{L}_{\hat{x}}(\bm{C})\mathcal{E}_R(\bm{A})\mathcal{L}_{\hat{x}}^\dagger(\bm{C})
    \label{inverse_E_1}
\end{gather}
\begin{gather}
    \nonumber
    E_{R,\hat{y}}(\bm{A})=(E_{\mathcal{L}})_{L,\hat{y}}(\bm{A})+\mathcal{E}_R(\bm{B}),\\
    E_{R,\hat{x}}(\bm{C})=(E_{\mathcal{L}})_{L,\hat{x}}(\bm{C})-\mathcal{L}_{\hat{x}}(\bm{C})\mathcal{E}_R(\bm{A})\mathcal{L}_{\hat{x}}^\dagger(\bm{C}),
    \label{inverse_E_2}
\end{gather}
\begin{gather}
    \nonumber
    E_{L,\hat{x}}(\bm{C})=-\mathcal{L}^\dagger_{\hat{y}}(\bm{A})\mathcal{E}_R(\bm{B})\mathcal{L}_{\hat{y}}(\bm{A})-\mathcal{E}_L(\bm{C})-(E_\mathcal{L})_{L,\hat{x}}(\bm{C}),\\
    E_{L,\hat{x}}(\bm{B})=-\mathcal{E}_R(\bm{C})-\mathcal{E}_L(\bm{B}),
    \label{inverse_E_3}
\end{gather}
\begin{align}
    \nonumber
    E_{L,\hat{y}}(\bm{A})=&-\mathcal{E}_R(\bm{C})-(E_{\mathcal{L}})_{L,\hat{y}}(\bm{A})-\mathcal{E}_L(\bm{A})\\
    &-\mathcal{E}_L(\bm{B})-\mathcal{E}_R(\bm{B}),
    \label{inverse_E_4}
\end{align}
as a function of the dual variables.

The same procedure can be repeated for larger tori, as exemplified in Fig. \ref{basic_CT + plaquette_changeofvars}(c). The net result, after the resolution of the Gauss law, is the same as for the minimal torus, i.e., a set of independent plaquettes and two large loops wrapping the lattice. We refer to Appendix \ref{generalization_square_lattice} for the complete derivation of the dual degrees of freedom.\\
%%%%%%%%%%%%%%%%%%%%%%%%%%%%%%%%%%%%%%%%%%%%%%%%%%%%%%%%%%%%%%%%%%%%%%%%%%%%
\section{\label{H_singleperiodicplaq_PBC}Loop Hamiltonian dynamics}
The dynamics of the single periodic plaquette is ruled by the KS Hamiltonian in Eq. \eqref{initial_KS_Hamiltonian}. We now express it in terms of the dual variables and obtain the encoded Hamiltonian that describes the loops dynamics. We discuss separately the features of the magnetic and electric contributions.

In the dual basis, the magnetic part of the Hamiltonian $H_B$ is the simplest, as $\text{Tr}\;U_P=\text{Tr}\;W_{\bm{n}}$ in terms of the plaquette operators, where the plaquettes $P$ are related to the sites $\bm{n}=\bm{A, B, C, D}$ as indicated in Fig. \ref{basic_CT + plaquette_changeofvars}(b). The independent loops are $W_{\bm{A},\bm{B},\bm{C}}$, while for the remaining site $\bm{n}=\bm{D}$ we have the relation
\begin{equation*}
    W_{\bm{D}}^\dagger=W_{\bm{A}}W_{\bm{B}}W_{\bm{C}}
\end{equation*}
due to the topology of the torus. The magnetic contribution is then
\begin{equation}
    H_B = \frac{1}{2g^2} \text{Tr}\bigg[4-\sum_{\bm{n}}W_{\bm{n}}-\prod_{\bm{n}}W_{\bm{n}}\bigg] + \text{H.c.},
    \label{reformulated_magnetic_H}
\end{equation}
where $\bm{n}$ runs over the independent plaquettes. All terms in the magnetic contribution turn out to be noninteracting, except for plaquette $W_{\bm{D}}$, as expected from the dualization procedure on the minimal torus \cite{Kaplan-Stryker}.

On the other hand, the electric term is now describing interactions and can be obtained by the squares of the inverse relations for the electric fields. In doing this operation, we take advantage of the property
\begin{equation*}
    \text{Tr}\;E^2_{\mu}(\bm{n})=\frac{1}{2}E^a_{L,\mu}(\bm{n})E^a_{L,\mu}(\bm{n})=\frac{1}{2}E^a_{R,\mu}(\bm{n}+\hat{\mu})E^a_{R,\mu}(\bm{n}+\hat{\mu}),
\end{equation*}
meaning that we can use either the left or right components of the electric fields to compute this Hamiltonian contribution. We can identify two different electric contributions, namely
\begin{equation*}
    H_E=H_{E,\text{loc}}+H_{E,\text{non-loc}},
\end{equation*}
where the subscripts stand for local and non-local, respectively. Their explicit expressions as a function of the loop electric fields are
\begin{widetext}
\begin{equation}
    H_{E,\text{loc}}=g^2[2\mathcal{E}_L^2(\bm{A})+2\mathcal{E}_L^2(\bm{C})+3\mathcal{E}_L^2(\bm{B})+(E_{\mathcal{L}})^2_{L,\hat{y}}(\bm{A})+(E_{\mathcal{L}})^2_{L,\hat{x}}(\bm{C})],
    \label{HE_local}
\end{equation}
\begin{align}
    \nonumber
    H_{E,\text{non-loc}}=g^2&[(E_{\mathcal{L}})_{L,\hat{x}}(\bm{C})\mathcal{E}_L(\bm{C})+(E_{\mathcal{L}})_{L,\hat{y}}(\bm{A})\mathcal{E}_L(\bm{A})+\mathcal{E}_L(\bm{C})\mathcal{R}[\mathcal{L}^\dagger_{\hat{x}}(\bm{C})]\mathcal{E}_R(\bm{A})\\
    \nonumber
    &+(E_{\mathcal{L}})_{R,\hat{x}}(\bm{C})\mathcal{E}_R(\bm{B})+\mathcal{E}_L(\bm{C})\mathcal{R}[\mathcal{L}_{\hat{y}}(\bm{A})]\mathcal{E}_R(\bm{B})\\
    \nonumber
    &+(E_{\mathcal{L}})_{L,\hat{x}}(\bm{C})\mathcal{R}[\mathcal{L}_{\hat{y}}(\bm{A})]\mathcal{E}_R(\bm{B})+2\mathcal{E}_R(\bm{C})\mathcal{E}_L(\bm{B})+\mathcal{E}_R(\bm{C})\mathcal{E}_L(\bm{A})\\
    \nonumber
    &+\mathcal{E}_R(\bm{C})\mathcal{E}_R(\bm{B})+\mathcal{E}_R(\bm{C})(E_{\mathcal{L}})_{L,\hat{y}}(\bm{A})+\mathcal{E}_L(\bm{A})\mathcal{E}_L(\bm{B})+\mathcal{E}_L(\bm{A})\mathcal{E}_R(\bm{B})\\
    &+\mathcal{E}_L(\bm{B})\mathcal{E}_R(\bm{B})+\mathcal{E}_L(\bm{B})(E_{\mathcal{L}})_{L,\hat{y}}(\bm{A})+\mathcal{E}_R(\bm{B})(E_{\mathcal{L}})_{L,\hat{y}}(\bm{A})].
    \label{HE_non-local}
\end{align}
\end{widetext}
The local contribution involves the squares of all the new independent loop electric fields, with coefficients related to the chosen sequence of CTs. The non-local terms instead are of the form
\begin{align*}
    \nonumber
    \mathcal{E}_{L,R}(\bm{N})\mathcal{P}\mathcal{E}_{L,R}(\bm{M})\mathcal{P}^\dagger&=\mathcal{E}_{L,R}(\bm{N})\mathcal{R}(\mathcal{P})\mathcal{E}_{L,R}(\bm{M})\\
    &\equiv\mathcal{E}^a_{L,R}(\bm{N})\mathcal{R}^{ab}(\mathcal{P})\mathcal{E}_{L,R}^b(\bm{M}),
\end{align*}
where the path $\mathcal{P}$ connects the site $\bm{N}$ to $\bm{M}$. After the explicit resolution of the Gauss laws, these non-local terms are expected, and connect electric fields in different points of the lattice through the parallel transports $\mathcal{R}(\mathcal{P})$. As for the coefficients of the local terms, the form of the paths $\mathcal{P}$ is not unique and is related to the set of employed CTs.

Due to the resolution of the Gauss laws according to the scheme of Fig. \ref{basic_CT + plaquette_changeofvars}(b), the dual Hamiltonian $H=H_B+H_{E,\text{loc}}+H_{E,\text{non-loc}}$ is not manifestly symmetric under the exchange of the two large loops $\mathcal{L}_{\hat{x}}$, $\mathcal{L}_{\hat{y}}$ and of the plaquette loops $W_{\bm{A}}$, $W_{\bm{B}}$. This is because the origin of the just mentioned loops is not placed symmetrically with respect to the diagonal line $\bm{A}\rightarrow\bm{C}\rightarrow\bm{A}$, as can be seen from the rightmost panel of Fig. \ref{basic_CT + plaquette_changeofvars}(b). Finally, we notice that $[H,(E_{\mathcal{L}})_{L,\hat{x}}(\bm{C})]\neq 0$ and $[H,(E_{\mathcal{L}})_{L,\hat{y}}(\bm{A})]\neq 0$ (see Appendix \ref{electric_loop_commutators} for the computational details), implying that the large loops are not constants of motion. This feature distinguishes non-Abelian $SU(N)$ LGTs from pure $U(1)$ Abelian LGT, where the large loops commute with the encoded Hamiltonian \cite{Haase2021resourceefficient}. In the latter case the commutation occurs because the gauge field of a commutative gauge group is not charged. In the present non-Abelian case, the gauge field itself is charged, leading to non-zero values for the commutators.

%%%%%%%%%%%%%%%%%%%%%%%%%%%%%%%%%%%%%%%%%%%%%%%%%%%%%%%%%%%%%%%%%%%%%%%%%%%%

\section{\label{work_basis_periodicplaqPBC}The case of $SU(2)$}
Even if our reformulation in terms of closed loops applies to $SU(N)$ LGTs, from now on we focus on the simplest of such groups, i.e., the rotation group $\mathcal{G}=SU(2)$. As we are going to show in the remainder of the paper, in this simple but still far from trivial case we can work out explicitly the group basis structure and deal with the single periodic plaquette problem in an efficient way.

We parametrize the elements of $SU(2)$ using the axis-angle coordinates, i.e., by specifying the rotation axis through a unit vector $\hat{n}(\theta,\phi)$ and the angle $\omega$ of rotation around it. This allows for the identification of the frame of coordinates ${\bf \Omega}=(\omega,\theta,\phi)$ in the 3D space. We refer to the references \cite{Khersonskii1988,sakurai1994modern,Bauer2023} for clear and pedagogical explanations of this and other representations of the rotation group. For the present purposes, we summarize the relevant properties of the axis-angle representation in Appendix \ref{SU2_grouptheory_recap}. 

Given the representation of the basis states as group elements, the first point we address is the identification of a suitable basis for the dual Hamiltonian, which we refer to as the local basis. On top of this, the main idea behind our approach is a variational determination of the basis coefficients to interpolate between the regions of the phase diagram dominated by electric and magnetic interactions, respectively, as pictorially depicted in Fig. \ref{TeaserVariational}(c-d). With this procedure, we are able to extract the running coupling efficiently for all values of the bare coupling constant.

\subsection{\label{localH_SU2}The local basis}
The dual Hamiltonian lends itself to be treated in the position representation, where the basis states are the eigenstates of the loop operators $\mathcal{L}\in\mathcal{W}$, rather than in the electric basis. As a further consideration, the primary limitation of the electric basis is its reduced efficiency in the weak coupling limit, where the magnetic Hamiltonian dominates, and the lowest-energy eigenstates involve sums over a high number of irreducible representations. We then consider $L^2(\mathcal{G},d\mu(g))$ as the Hilbert space of a single loop, that is, the square integrable functions over the Lie group $SU(2)$ with respect to the invariant Haar measure $d\mu(g)$. Employing the axis-angle coordinates to parametrize group elements (see Appendix \ref{SU2_grouptheory_recap}), a general state of the system $|\psi({\bf \Omega})\rangle$ is written as the superposition
\begin{equation*}
    |\psi({\bf\Omega})\rangle=\int\;d\mu({\bf\Omega})\;\psi({\bf\Omega})|g({\bf\Omega})\rangle,
\end{equation*}
with orthogonal basis states $|g\rangle\in SU(2)$, satisfying the relation
\begin{equation}
    \langle \psi({\bf\Omega})|\psi'({\bf\Omega})\rangle=\int d\mu({\bf\Omega})\;\psi^*({\bf\Omega})\psi'({\bf\Omega}).
\end{equation}

As a first step, we identify the proper basis for the reformulated Hamiltonian as composed by the eigenstates of the local Hamiltonian 
\begin{equation}
    H_{\text{loc}}=\frac{1}{2g^2}\sum_{\bm{n}\neq\bm{D}}\text{Tr}[2-(W_{\bm{n}}+W^\dagger_{\bm{n}})]+H_{E,\text{loc}}
    \label{local_H}
\end{equation}
in the group basis. The local Hamiltonian can be written as sum of single plaquette Hamiltonians
\begin{equation*}
    H_{0,\bm{n}}(g_i)\equiv \frac{1}{2g_i^2}\text{Tr}[2-(W_{\bm{n}}+W^\dagger_{\bm{n}})]+2g_i^2\mathcal{E}^2(\bm{n})
\end{equation*}
for $\bm{n}=\bm{A},\bm{B},\bm{C}$, plus the electric contributions of the two large loops, without the associated magnetic terms. We notice that, as suggested from Eq. \eqref{HE_local}, the coefficients of the various $H_{0,\bm{n}}(g_i)$ depend on the loop $W_{\bm{n}}$, and are generically different from each other \footnote{A change in the electric coefficient $g^2\rightarrow\alpha g^2$ of the KS Hamiltonian means that we can study the model of Eq. \eqref{initial_KS_Hamiltonian} with a rescaled coupling $\tilde{g}=\sqrt{\alpha}g$, up to an overall multiplicative constant $\sqrt{\alpha}$.}. 

In the axis-angle coordinates, the single plaquette Hamiltonian has the form
\begin{align}
    \nonumber
    H_{0,\bm{n}}(g_i)&=\frac{2}{g_i^2}\bigg(1-\cos\frac{\omega}{2}\bigg)\\
    &+2g_i^2\bigg[\frac{{\bf L}^2}{\sin^2\frac{\omega}{2}}-4\bigg(\frac{\partial^2}{\partial\omega^2}+\cot\frac{\omega}{2}\frac{\partial}{\partial\omega}\bigg)\bigg].
\end{align}
Since the square loop electric field operator contains ${\bf L}^2$, it becomes very useful to expand the $(\theta,\phi)$-angular part in terms of spherical harmonics, using the mixed basis expansion introduced in Ref. \cite{Bauer2023}
\begin{gather}
    \nonumber
    \psi({\bf \Omega})=\mathcal{N}(\omega)\sum_{\ell,m}c_{\ell,m}Y_{\ell,m}(\theta,\phi)u_{\ell,m}(\omega),\\
    \mathcal{N}(\omega)=\frac{1}{2\sin\frac{\omega}{2}}
    \label{basis_exp_Bauer}
\end{gather}
where $\mathcal{N}(\omega)$ is a factor to ensure the proper normalization of the wave function $\psi({\bf \Omega})$, coefficients $c_{\ell,m}$ and reduced function $u_{\ell,m}(\omega)$ at fixed values of $\ell,m$, with respect to the group measure. Within this representation, the single plaquette eigenproblem $H_{0,\bm{n}}(g_i)|\psi({\bf\Omega})\rangle=\epsilon|\psi({\bf\Omega})\rangle$ becomes a differential equation in the angular coordinate $\omega$. We introduce the reduced wave function
\begin{equation*}
    u_{\ell,m}(\omega)=R_{\ell,m}(\omega)\mathcal{N}^{-1}(\omega)=2R_{\ell,m}(\omega)\sin\bigg(\frac{\omega}{2}\bigg)
\end{equation*}
dictated by Eq. \eqref{basis_exp_Bauer}, allowing for the rewriting of the Schroedinger equation as
\begin{gather}
    \nonumber
    -u_{\ell,m}''(\omega)+\frac{1}{4}\bigg[\frac{\ell(\ell+1)}{\sin^2\omega/2}+\frac{1}{g^4}\bigg(1-\cos\frac{\omega}{2}\bigg)-1\bigg]u_{\ell,m}(\omega)\\
    =\tilde{\epsilon}u_{\ell,m}(\omega),\qquad\tilde{\epsilon}\equiv\frac{\epsilon}{8g^2}.
    \label{diffeq_reduced_wf_omega}
\end{gather}
This is a central motion problem on the 3D sphere, with central potential given by the magnetic Hamiltonian. At fixed value of $\ell$, the eigenfunction $u_{\ell,m}(\omega)$ has degeneracy $2\ell+1$, due to the azimuthal quantum number $m$. To label the eigenstates of Eq. \eqref{diffeq_reduced_wf_omega}, we introduce the ket $|\alpha_{\ell},\ell,m\rangle$. This involves the angular momentum quantum numbers $\ell,m$ and an integer index $\alpha_\ell$ denoting the state of the reduced wave function $u_\alpha(\omega)$ at fixed $\ell$. For example, $\alpha_0=0$ denotes the ground state of $\ell=0$, $\alpha_0=1$ the first excited state, and so on and so forth. 

By taking into account that Eq. \eqref{local_H} can be written as the sum
\begin{align}
    \nonumber
    H_{\text{loc}}=&H_{0,\bm{A}}(g)+H_{0,\bm{B}}(\sqrt{3/2}g)+H_{0,\bm{C}}(g)\\
    &+g^2[(E_{\mathcal{L}})^2_{L,\hat{y}}(\bm{A})+(E_{\mathcal{L}})^2_{L,\hat{x}}(\bm{C})],
    \label{local_H_sums_plaquettes_loops}
\end{align}
the local basis for the minimal torus is the tensor product of the basis obtained through the solutions of Eq. \eqref{diffeq_reduced_wf_omega} for the various contributions to Eq. \eqref{local_H_sums_plaquettes_loops}. We note that the factor of $\sqrt{3/2}$ comes from the specific CTs employed in the dualization procedure. Regarding the large loops, from Eq. \eqref{local_H_sums_plaquettes_loops} it is clear that their contributions to $H_{\text{loc}}$ can be also interpreted as $H_{0,\mathcal{L}}(g\rightarrow\infty)$, i.e., they do not have the magnetic contribution. Consequently, the right basis to describe them locally would be the electric one. While previous studies have explored the digitization of the link degree of freedom through the discretization of the group coordinate $\omega$ \cite{Bauer2023}, following previous works on $U(1)$ \cite{Haase2021resourceefficient,BauerPRD2023}, we instead use the eigenbasis of the local Hamiltonian. Since this eigenbasis and the related matrix elements can be efficiently computed classically, the required resources for both classical and quantum computation are significantly reduced.

To lighten the notation, we introduce more compact labels for the mixed basis states, using a Fock enumeration of the tensor product states. In the following, loops are ordered as $(W_{\bm{A}},W_{\bm{B}},W_{\bm{C}},\mathcal{L}_{\hat{x}},\mathcal{L}_{\hat{y}})$. By denoting the ket associated to each local basis state with $|\alpha_{\ell_i},\ell_i,m_i\rangle_{g_i}$, where we added a subscript for the coupling $g_i$, we write the tensor product explicitly as
\begin{equation}
    \bigotimes_{i=1}^5|\alpha_{\ell_i},\ell_i,m_i\rangle_{g_i}\equiv\ket{I_1(g_1),I_2(g_2),I_3(g_3),I_4(g_4),I_5(g_5)},
    \label{local_basis_H0_mixedangle}
\end{equation}
where $I_k(g_k)$ is an integer number labeling the state of the $k$-th loop variable at value $g_k$ of the local coupling in Eq. \eqref{local_H}, and is related to the mixed-basis quantum number as
\begin{equation}
    \ket{I(g_k)}=\ket{\alpha_{\ell_k},\ell_k,m_k}_{g_k}.
    \label{general_Fock_basis_AM_mapping}
\end{equation}
The mapping between the integer $I$ and the mixed basis quantum numbers takes into account the angular momentum degeneracy, and enumerates the states in ascending order in $\tilde{\epsilon}$. To further clarify the notation, let us consider the example case of $L_{max}=5$ states for each loop variable, i.e., values of $I\in\{0,\ldots,4\}$. For the local basis, we write
\begin{equation}
    \begin{Bmatrix}
        \ket{0}\\
        \ket{1}\\
        \ket{2}\\
        \ket{3}\\
        \ket{4}
    \end{Bmatrix}
    \quad
    \leftrightarrow
    \quad
    \begin{Bmatrix}
        \ket{0_0,0,0}\\
        \ket{0_1,1,-1}\\
        \ket{0_1,1,0}\\
        \ket{0_1,1,1}\\
        \ket{1_0,0,0}
    \end{Bmatrix}
    \label{Fock_basis_mapping_5states}
\end{equation}
and the full basis has size $L_{max}^5=5^5$. Using this labeling, it is easier to reconstruct the quantum numbers of a given basis element. For example, $\ket{0,2,1,2,4}$ is a short notation for
\begin{equation}
   \ket{0_0,0,0}\otimes\ket{0_1,1,0}\otimes\ket{0_1,1,-1}\otimes\ket{0_1,1,0}\otimes\ket{1_0,0,0}.
\end{equation}

%%%%%%%%%%%%%%%%%%%%%%%%%%%%%%%%%%%%%%%%%%%%%%%%%%%%%%%%%%%%%%%%%%%%%%%%%%%%

\subsection{\label{non-loc_H_matrix_elems}The non-local Hamiltonian matrix elements}
After identifying the local basis, which is physically determined by the single plaquette problem, we need to compute the matrix elements of the full Hamiltonian derived from our resource-efficient reformulation. In addition to the local terms discussed in the current Section, the primary challenge lies in the non-local terms present in both the magnetic and electric Hamiltonians. 

In this Subsection, we briefly analyze the action of the left loop electric field and loop operators on states written in mixed-angle representation. Given that we can express both of these operators in terms of scalars and vectors under rotations generated by the orbital angular momentum ${\bf{L}}$, we can apply the Wigner--Eckart theorem to compute a generic matrix element involving them \cite{Bauer2023}. This theorem states that, for any spherical tensor $T_k$ of rank $k$, general matrix elements can be written as
\begin{align}
    \nonumber
    \bra{\{\alpha'\},\ell',m'}T^{(q)}_k\ket{\{\alpha\},\ell,m}=&\langle\{\alpha'\},\ell'||T_k||\{\alpha\},\ell\rangle\\
    &\cdot\braket{\ell'm'|\ell m,kq},
    \label{W-E_theorem}
\end{align}
where $\{\alpha,\alpha'\}$ are non-rotational quantum numbers, and the superscript $q$ labels the component of the spherical tensor \cite{sakurai1994modern}. In the case of $H_{\text{non-loc}}$, only scalar or vector operators are involved ($k=0,1$) and the only additional quantum number we have to consider is the label $\alpha_\ell$, specifying the reduced wave function $u_{\alpha_\ell}(\omega)$. The computational details of the matrix elements are reported in Appendix \ref{WE_electric_field_computations}, \ref{WE_EL_ER_product_computations}, \ref{WE_loop_computations} and \ref{Rab_scalar_vector_decomposition}.

In the remainder of this Section, we numerically compute the matrix element of the dual Hamiltonian for the minimal torus in the local basis outlined in Eq. \eqref{local_basis_H0_mixedangle}. However, since this basis has infinite elements, we need to truncate it and retain only a finite number of states for each loop. If we keep $L_{max}$ states for each $H_0(g_i)$, by taking into account also the degeneracy of $2\ell+1$ at fixed orbital quantum number $\ell$, the square matrix associated to the minimal torus has size $L_{max}^5\times L_{max}^5$.

%%%%%%%%%%%%%%%%%%%%%%%%%%%%%%%%%%%%%%%%%%%%%%%%%%%%%%%%%%%%%%%%%%%%%%%%%%%%

\subsection{\label{variational_GS}The variational principle and optimal choice of truncated local basis}
\begin{figure*}[t!]
    \centering
    \includegraphics[width=0.9\linewidth]{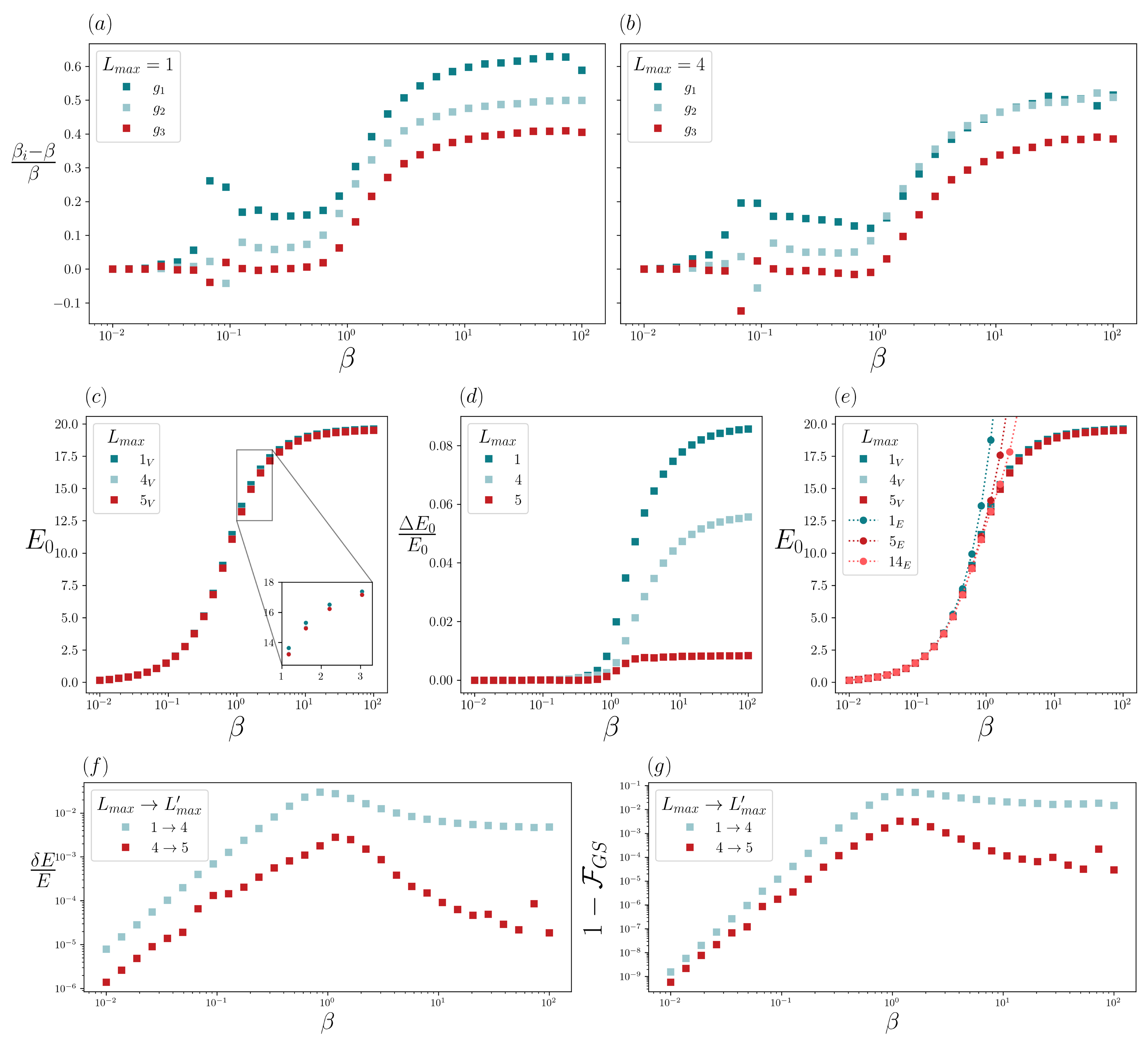}
    \caption{(a-b) Plot of the distance between local couplings $(\beta_1,\beta_2,\beta_3)=(1/2g_1^2,3/4g_2^2,1/2g_3^2)$, determined through variational optimization, and the bare coupling $\beta=(2g^2)^{-1}$, as a function of the bare coupling $\beta$ for the different values of truncations $L_{max}=1,4$ (left and right subplots, respectively). Instabilities in the data points are associated to the numerical minimization. (c) Energies of the variationally determined ground states as a function of the bare coupling $\beta$ for the different values of truncations $L_{max}=1,4,5$. (d) Relative energy difference $\Delta E_0/E_0=(E_0(\bm{g_0})-E_0(\bm{g_V}))/E_0(\bm{g_V})$ between the initial ansatz $\bm{g}_0$ and the optimal couplings as a function of the bare coupling $\beta$ for the different values of truncations $L_{max}=1,4,5$. The data shows the absence of energy reduction for strong couplings, which explains the numerical instabilities of panels (a-b) for $\beta\lesssim 1$. (e) Energies of both variationally determined ground states and ground states in the electric representation as a function of the bare coupling $\beta$ for the different values of truncations $L_{max}=1,4,5$. We observe that the electric representation is efficient in the strong coupling regime, while it is increasingly worse towards weak coupling. In contrast, our approach efficiently interpolates between strong and weak coupling regimes. (f) Relative energy differences $\delta E/E$ between consecutive values of the truncation $L_{max}$ as a function of the bare coupling $\beta$. (g) Infidelities $ 1-\mathcal{F}_{\text{GS}}$ between consecutive values of the truncation $L_{max}$ as a function of the bare coupling $\beta$.}
    \label{joint_energy_beta_etc}
\end{figure*}
In the previous Section \ref{localH_SU2}, we discussed the truncation of the gauge degrees of freedom according to the local Hamiltonian described by Eq. \eqref{local_H}. In particular, the explicit expression in Eq. \eqref{local_H_sums_plaquettes_loops} suggests the values of the couplings $g_i$ to be used for each plaquette, thereby determining the local eigenstates and, consequently, the Hilbert space to which we truncate. It is important to point out that the ground state obtained in this way is only the state of minimum energy within the truncated space. While this approach represents a significant improvement over other truncation schemes, such as the electric \cite{ZoharPRD2015,ByrnesPRA2006} or magnetic representations \cite{LammPRD2020,DiscreteLammPRD2022}, further enhancement can be achieved by the \textit{variational optimization} of the couplings ${g_i}$. Such optimization is motivated by the presence of non-local terms in the Hamiltonian, as evident from both the magnetic and electric contributions given by Eqs. \eqref{reformulated_magnetic_H} and \eqref{HE_non-local}, respectively.

In search of the optimized ansatz, we adopt the following procedure. First, the initial values for the local couplings ${g_i}$ are taken to be those suggested by the reformulated local Hamiltonian, namely $(g_1,g_2,g_3,g_4,g_5)=(g,\sqrt{\frac{3}{2}}g,g,\infty,\infty)\equiv \bm{g}_0$, as detailed in Eq. \eqref{local_H_sums_plaquettes_loops}. We then compute variationally the values of the local coupling to get the optimal local basis $\ket{I_1(g_1),\ldots,I_5(g_5)}\equiv\ket{\bm{I}(\bm{g}_I)}$, for \textit{any} value of the bare coupling $g$. This amounts to evaluate the energy functional
\begin{equation}
    E(\beta,\bm{g}_I)=\bra{\Psi(\bm{g}_I)}H_{\text{dual}}\ket{\Psi(\bm{g}_I)}
    \label{energy_functional_paramspace}
\end{equation}
in the parameter space of the local couplings, where we introduced $\beta\equiv(2g^2)^{-1}$ to write the bare coupling and avoid confusion with the local couplings $g_i$. As just stated, the initial ansatz for the ground state $\Psi(\bm{g}_I)$ is evaluated in the corresponding initial values $\bm{g}_0$. The ground state energy of the system is estimated as
\begin{equation}
    E_0(\beta)=\min_{\bm{g}_I}E(\beta,\bm{g}_I),
    \label{GS_minimization_energy_functional_paramspace}
\end{equation}
where the minimization is performed by keeping $L_{max}$ states for each loop. Thanks to the variational principle \cite{sakurai1994modern}, with this computation we obtain an upper bound estimate for the ground state energy, which turns out to be exact in the limit $L_{max}\rightarrow\infty$, when the complete local basis is considered without truncation. As a first simplification and working hypothesis, in all the cases considered in this Section we retain one state for the large loops and $L_{max}$ states for each plaquette \footnote{Contrary to the Abelian case, the large loops do not decouple in the non-Abelian theory even in the absence of dynamical charges, because the gauge field itself is charged. However, as we verified numerically, the large loops remain almost decoupled and can be well described by a single state in the explored region of couplings. On physical ground, we may expect this behavior due to the absence of a bare plaquette term for the large loops Hamiltonian.}. 

For any considered value of $\beta\in[10^{-2},10^2]$, we apply this scheme in two steps: in the first step, only the plaquette local couplings $(g_1,g_2,g_3)$ are left free, while those of the large loops are fixed at the initial values $(g_4,g_5)=(\infty,\infty)$. The minimization process iteratively determines the optimal couplings $(g_1,g_2,g_3)$ through Eq. \eqref{GS_minimization_energy_functional_paramspace}; in the second step, the plaquette local couplings are held fixed at the optimal values while the large loops are variationally optimized. Both iterations continue until the required precision in the energy and couplings is achieved (see Appendix \ref{numerical_implementation} for more details). 

We plot in Fig. \ref{joint_energy_beta_etc}(a-b) the results of the first optimization step. We show the relative displacement of the values of $(\beta_1,\beta_2,\beta_3)=(1/2g_1^2,3/4g_2^2,1/2g_3^2)$ determined through the variational procedure for the local coupling of the plaquettes with respect to their bare values. We observe that in the strong coupling regime, when $\beta\rightarrow 0$, the deviations from the initial ansatz of the local couplings are small, while in the weak coupling limit $\beta\rightarrow\infty$ the variational basis is different from the one suggested from the dual formulation. This observation is independent from the chosen truncation $L_{max}$, and signals the renormalization of the local couplings with respect to the bare ones as long as we approach the weak coupling limit. 

Alongside this, we plot the ground state energies $E_0$ as a function of $\beta$ in Fig. \ref{joint_energy_beta_etc}(c). Consistently with the previous observation, we observe no real difference between the initial ansatz $\bm{g}_0$ and the optimal couplings in the strong coupling regime, while in the weak coupling regime the ground state is lower. This behavior is even more evident in Fig. \ref{joint_energy_beta_etc}(d), where we show the relative energy difference
\begin{equation}
    \frac{\Delta E_0}{E_0}\equiv\frac{E_0(\bm{g_0})-E_0(\bm{g_V})}{E_0(\bm{g_V})}
    \label{relative_diffE}
\end{equation}
as a function of $\beta$. The absence of energy reduction at strong coupling explains the numerical instability found in Fig. \ref{joint_energy_beta_etc} (a-b) for $\beta\lesssim 1$. Indeed, even if in that region the variational optimization produces a significant exploration of parameter space, this only leads to a negligible lowering of the relative energy difference. We then compare, in Fig. \ref{joint_energy_beta_etc}(e), the ground state energies obtained with variational optimization to those in the electric representation, with the same truncation $L_{max}$. Our variational protocol effectively interpolates between the strong and weak coupling regimes. As a common feature for the different truncations, the relative energy difference is very small deep in the strong coupling limit, where $\beta\approx 10^{-2}$, and increase as long as we go towards the weak coupling regime, with a maximum for $\beta\approx 1$. Around this value, we have the maximal competition between the electric and magnetic Hamiltonians.

As a final comment, according to the variational principle, when the truncation parameter $L_{max}$ is increased the ground state energy is lowered towards its exact value, obtaining a strictly decreasing sequence for any value of the bare coupling. In Fig. \ref{joint_energy_beta_etc}(f) we show the energy difference
\begin{equation}
    \delta E=|E_{0,L_{max}}(\beta,g_i)-E_{0,L_{max}'}(\beta,g_i)|,
\end{equation}
for two consecutive values of the truncations $L_{max},\;L'_{max}$, to have indications about the energy precisions at every step of the variational procedure. At the same time we check the infidelity of the ground state wave function, defined as
\begin{equation}
    1-\mathcal{F}_{\text{GS}}=1-|\langle\psi_{0,L_{max}}(\beta,g_i)|\psi_{0,L'_{max}}(\beta,g_i)\rangle|^2
\end{equation}
with the same convention on the truncation values, and plot it in Fig. \ref{joint_energy_beta_etc}(g). As expected, the infidelity is smaller if a larger number of states is included, and peaks around $\beta\approx O(1)$, i.e., the value above which the plaquette local basis is significantly more efficient with respect to the electric one.

We now comment on the second step of our procedure, corresponding to the variational determination of the large loops local couplings, once the plaquette ones are fixed to their optimal values. In the explored region of the coupling, $\beta\in[10^{-2},10^2]$, we find that the ground-state energy displays a plateau for $g_{4,5}\gg g$ when only one state per large loop is retained. We conclude that, for practical purposes, it is convenient to disregard the second iteration step and keep $g_{4,5}$ constant and large.    

%%%%%%%%%%%%%%%%%%%%%%%%%%%%%%%%%%%%%%%%%%%%%%%%%%%%%%%%%%%%%%%%%%%%%%%%%%%%
\subsection{\label{running_coupling}Running coupling computation}
\begin{figure*}[t!]
    \centering
    \includegraphics[width=0.9\linewidth]{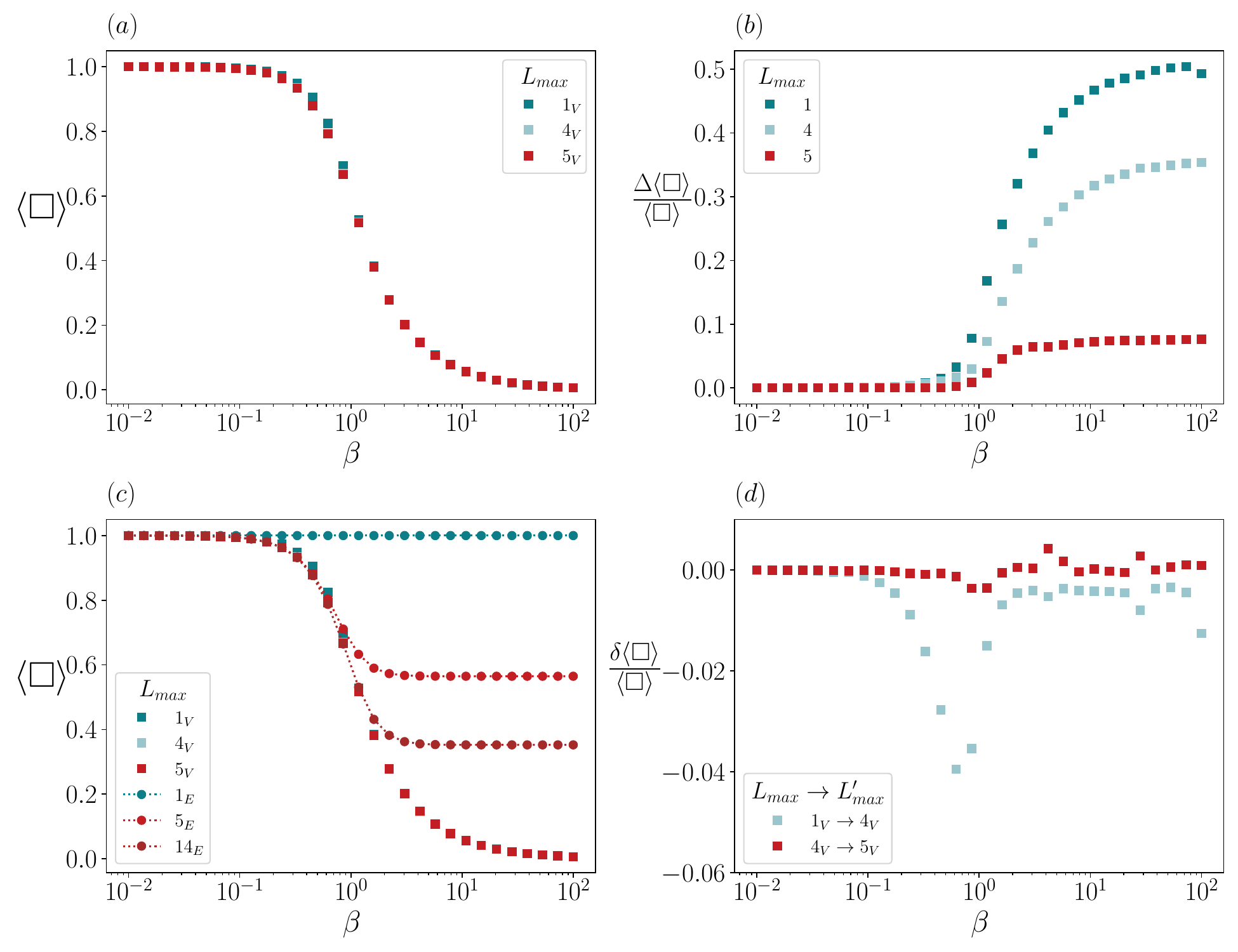}
    \caption{(a) Expectation values of the plaquette operator $ \langle\square\rangle$ on the variationally determined ground states as a function of the bare coupling $\beta$, for the different values of truncations $L_{max}=1,4,5$. (b) Relative difference $\frac{\Delta \langle\square\rangle}{\langle\square\rangle}\equiv\frac{\langle\square\rangle(\bm{g_0})-\langle\square\rangle(\bm{g_V})}{\langle\square\rangle(\bm{g_V})}$ in the expectation values of the plaquette operator between the initial ansatz $\bm{g}_0$ and the optimal couplings as a function of the bare coupling $\beta$, for the different values of truncations $L_{max}=1,4,5$. As observed for the energy (see Fig. \ref{joint_energy_beta_etc}), the variational approach is more advantageous for smaller $L_{max}$. (c) Expectation values of the plaquette operator $ \langle\square\rangle$ on both the variationally determined ground states and ground states in the electric representation as a function of the bare coupling $\beta$, for the different values of truncations $L_{max}=1,4,5$. (d) Relative differences in the expectation value of the plaquette operator $\delta \langle\square\rangle/\langle\square\rangle\equiv(\langle\square\rangle_{L_{max}'}-\langle\square\rangle_{L_{max}})/\langle \square\rangle_{L_{max}'}$ between consecutive values of the truncation $L_{max}$ as a function of the bare coupling $\beta$.}
    \label{running_couplings_plot}
\end{figure*}
In this Section we estimate the expectation value of the plaquette operator, i.e., the local observable related to the magnetic contribution $H_B$, which is a key quantity in the study of LGTs. Indeed, its dependence on $g^{-2}$ can be related to the running of the coupling \cite{BOOTH2001,Haase2021resourceefficient,PaulsonPRX2021,ClementePRD2022}. This last quantity refers to the dependence of the coupling constant on the specific energy scale at which is probed, and arises from quantum fluctuations effects. Its complete understanding is of absolute importance for several aspects, from the interpretation of experimental results of collider experiments to the theoretical understanding of fundamental interactions. However, due to limitations of Monte Carlo methods in the weak coupling limit \cite{Schaefer2011, Rothe}, the extrapolation of the running coupling towards the continuum limit stands as an open problem in LGTs.

Given the ground state $|\psi_0\rangle$ obtained with the variational procedure of Section \ref{variational_GS}, we compute the expectation value 
\begin{equation}
    \langle\square\rangle\equiv\frac{g^2}{2N_{\text{plaq}}}\langle\psi_0|H_B|\psi_0\rangle,
    \label{running_plaq_def}
\end{equation}
where $N_{\text{plaq}}$ is the number of plaquettes.

We plot in Fig. \ref{running_couplings_plot}(a) the results obtained for different values of truncations $L_{max}$. We observe that, even for the smallest truncation, the computation with the variationally optimized ground states leads to better precision at all values of the bare couplings $\beta$. In particular, the first increase in truncation size, i.e., $L_{max}=1\rightarrow L_{max}'=4$, leads to a relative difference $(\langle\square\rangle_{L_{max}'}-\langle\square\rangle_{L_{max}})/\langle \square\rangle_{L_{max}'}$ at most of 5\% in the running coupling, which is further reduced to 1\% when increasing the truncation size from $L_{max}=4\rightarrow L_{max}'=5$, see  Fig. \ref{running_couplings_plot}(d). The comparison with the initial ansatz for the variational procedure, i.e., the non-optimized dual basis of Eq. \eqref{local_H_sums_plaquettes_loops}, is showed in Fig. \ref{running_couplings_plot}(b), and highlights the improvement in the region around $\beta\approx O(1)$ when increasingly more states are included in the local basis. This behavior is a further sign that our variational procedure is interpolating between the strong (electric) and weak (magnetic) regimes of the bare couplings. Lastly, we compare in Fig. \ref{running_couplings_plot}(c) the variational computation of $\langle\square\rangle$ with the results obtained using the electric basis, showing definitely that this interpolation is done efficiently. Given the bounded nature of $\langle\square\rangle$ we provide further discussion of the plaquette operator in Appendix \ref{relative_plaquette}.

%%%%%%%%%%%%%%%%%%%%%%%%%%%%%%%%%%%%%%%%%%%%%%%%%%%%%%%%%%%%%%%%%%%%%%%%%%%%

\section{\label{conclusions}Conclusions and outlook}
In this paper, we have developed an efficient scheme for the classical and quantum simulation of non-Abelian LGTs beyond 1D that works in all regimes of the couplings and allows to approach the continuum limit with finite resources. The key ingredients of our scheme are:
\begin{itemize}
    \item the derivation of an encoded Hamiltonian for the gauge invariant degrees of freedom, small (plaquettes) and large loops, on periodic lattices;\\
    \item the semi-analytical variational identification of an optimal {\it coupling-dependent} basis for the loops to minimize the truncation error.
\end{itemize}
We have demonstrated the effectiveness of our scheme by determining the encoded Hamiltonian of $SU(N)$ gauge theories without charges on 2D arbitrary tori and the running of the coupling for $SU(2)$ gauge theory on a minimal torus. With only five states per plaquette, we have achieved a percent-level precision in computing both the ground state and the average value of the plaquette operator $\langle\square\rangle$ for all the values of the bare coupling $g$. 

Our scheme enables a proof-of-principle experimental demonstration of continuum limit computations for non-Abelian gauge theories on current quantum hardware \cite{Meth2025} and opens the door to several promising directions. One can use $\langle\square \rangle(g)$ as the physical observable to define the renormalized coupling constant \cite{CreutzSU(2)1980} $g_R$ at a given lattice spacing (subtraction scale) $a_0$. 
As outlined in \cite{ClementePRD2022}, the function $g=g(g_R,a)$ can be determined using a {\it step scaling approach} that combines Hamiltonian computations -- feasible at very weak coupling -- with large-scale Monte Carlo simulations. Recent results for pure $U(1)$ gauge theory \cite{crippa2024determining21dimensionalquantumelectrodynamics} suggest that meaningful matching can be achieved for $6\times 6$ lattices with open boundaries. The application of our scheme can potentially reduce the lattice size, due to the faster convergence with PBC, and it performs better even with lower truncations, thanks to the applied variational algorithm for the basis choice. Our construction is also applicable to the $U(1)$ gauge theory: it generalizes previous results \cite{Haase2021resourceefficient,BauerPRD2023} and can, in fact, improve them, as indicated by preliminary calculations using exact diagonalization. We plan to complete this comparison and also to study the performance with variational quantum circuits based on qudit architectures using Rydberg atoms \cite{GlaetzlePRX2014,DanielGCPRL2022} or trapped ions \cite{Meth2025}. Using qudits is indeed the most natural choice to represent finite truncations of higher-dimensional gauge fields, and the current state-of-the-art high fidelities of qudit gates allow for concrete proposals on the aforementioned platforms. Furthermore, in our proposal, the preparation of the ground state could be practically realized through a variational quantum eigensolver \cite{Peruzzo2014,McClean2016}, with additional variational parameters related to the local basis optimizations. In this respect, the simultaneous variational optimization of both the state (circuit) and the dual Hamiltonian (basis choice) can be performed without additional overhead, using the same measurements budget \cite{Kokail2019}.

%Possible future directions: tensor network based computations
Another important application regards tensor network computations. Combining this methodology with the proposed resource efficient formulation has the potential to optimize the computational efficiency and accuracy of classical simulations in gauge theories. In this context, the local coupling dependent variational ansatz for the ground state can also be used to develop quantum-inspired classical spin networks for Abelian and non-Abelian LGTs, enabling the computation of expectation values of observables towards the continuum limit, such as the running coupling or the shear viscosity \cite{Turro2024Feb}, for arbitrary tori of small sizes. To this end, it is important to understand how the non-local electric terms in the dual Hamiltonian behave as a function of the distance. To be more concrete, one should study the long-distance behavior of the ratio between the norm $||H_{E,\text{non-loc}}||$ and the first energy gap, which is of order $g^2$. If the ratio decays to a finite value, the long-range tails can be neglected and they can be treated as next-to-nearest neighbor tunnelings in the corresponding spin network. While this truncation becomes exact in the limit $g^2\rightarrow 0$ \cite{Mathur_PLB2015}, a general proof of this statement is, to the best of our knowledge, missing in the literature, and will be object of future in-depth study. Characterizing the scaling behavior of this ratio is crucial to address the more theoretical question of the density of entanglement needed to accurately describe the corresponding continuum field theory. Regarding the scalability, a general formulation of the local basis problem in terms of the eigenstates of the reduced density matrix of the complete (not truncated) problem can be done \cite{Materia2024Apr}, but it would be rather unpractical as it requires the knowledge of the full basis.

%Outlooks and completions: final discussion (most general local unitary which is hard, while we do something semi analytic + 3D)
Finally, even if we  performed numerical computations in the case of $SU(2)$ in two spatial dimensions, our method is general and can be applied to any $SU(N)$ LGT, in particular to $SU(3)$ LGT in three spatial dimensions. Indeed, the sequence of canonical transformations can be performed in the same way, with differences coming from the counting of fluxes when periodic boundary conditions are imposed, changing the number of independent closed surfaces \cite{Kaplan-Stryker}. The construction of the local basis always reduces to a $(N-1)$-dimensional differential equation, associated to a motion in a potential for any $N$. Another important extension is the inclusion of matter, which will be object of future works. A first target is represented by $SU(2)$ LGTs with dynamical charges in two dimensions, whose minimal realization requires resources already available \cite{Meth2025}. 

%%%%%%%%%%%%%%%%%%%%%%%%%%%%%%%%%%%%%%%%%%%%%%%%%%%%%%%%%%%%%%%%%%%%%%%%%%%%

\section{Acknowledgements}
We are very grateful to P. Calabrese, F. M. Surace and L. Tarruell for a critical reading of the manuscript. P.F. gratefully acknowledges A. Trombettoni and J. C. Pinto Barros for useful discussions and collaboration on a related topic. We acknowledge funding from MCIN/AEI/10.13039/501100011033 (LIGAS PID2020-112687GB-C22) and Generalitat de Catalunya (AGAUR 2021 SGR 00138). A.C. acknowledges support from the UAB Talent Research program. M.M. acknowledges support from MICINN/10.13039/501100004837 through an FPU grant (FPU22/03907). All authors acknowledge funding from EU QUANTERA DYNAMITE (funded by MICN/AEI/10.13039/501100011033 and by the European Union NextGenerationEU/PRTR PCI2022-132919 (Grant No. 101017733)), from Generalitat de Catalunya (QuantumCAT U16-011424, co-funded by ERDF Operational Program of Catalonia 2014-2020), from the Ministry of Economic Affairs and Digital Transformation of the Spanish Government through the QUANTUM ENIA project call Quantum Spain project. P.F. acknowledges support of the program Investigo (ref. 200076ID6/BDNS 664047), funded by the European Union through the Recovery, Transformation and Resilience Plan NextGenerationEU.

\appendix
\widetext 

%%%%%%%%%%%%%%%%%%%%%%%%%%%%%%%%%%%%%%%%%%%%%%%%%%%%%%%%%%%%%%%%%%%%%%%%%%%%

\section{\label{appendix_YM_action}$SU(N)$ continuum Yang-Mills action}
In this work we consider lattice field theories with local symmetry associated to the compact Lie group $SU(N)$. Its corresponding Lie algebra, $\mathfrak{su}(N)$, is spanned by a set of $N^2-1$ generators $T^a$, and we choose to work in the defining representation of the group, such that
\begin{equation}
    [T^a,T^b]=if^{abc}T^c,\qquad\qquad \text{Tr}(T^aT^b)=\frac{1}{2}\delta_{ab},
    \label{SU(N)_algebra_normalization}
\end{equation}
where $f^{abc}$ are the structure constants of the group, completely anti-symmetric in the internal indices $a,\;b,\;c$, and we fixed the normalization of the generators according to the standard convention \cite{Peskin,Scwhartz,Maggiore}. When $N=2$ we can identify the generators with the Pauli matrices $T^a=\sigma^a/2$, while for $N=3$ they are the Gell-Mann matrices $T^a=\lambda^a/2$.

In the continuum, the gauge potential $A_\mu=A_\mu^aT^a$ transforms as 
\begin{equation}
    A_\mu'(x)=\Omega(x)A_\mu(x)\Omega(x)^{-1}-i[\partial_\mu\Omega(x)]\Omega(x)^{-1},
\end{equation}
where $\Omega(x)\in SU(N)$ is a local gauge transformation. We also introduce the field strength tensor, defined in terms of the gauge potential as
\begin{equation}
    F_{\mu\nu}(A)=\partial_\mu A_\nu-\partial_\nu A_\mu-i[A_\mu,A_\nu],
\end{equation}
which is gauge covariant, i.e., such that $F_{\mu\nu}(A')=\Omega(x)F_{\mu\nu}(A)\Omega(x)^{-1}$. In the absence of matter, we can describe the dynamics of the gauge theory by means of the Yang-Mills (YM) action \cite{YangPR1954}
\begin{equation}
    S=-\frac{1}{2g^2}\int\;d{\bf x}\;\text{Tr}(F_{\mu\nu}F^{\mu\nu})
    \label{continuumYM_action}
\end{equation}
alongside the Gauss law constraint, which enforces local gauge invariance. The YM action can be regularized on the lattice in two ways: the first one involves the discretization of the continuum Lagrangian in Eq. \eqref{continuumYM_action}, and goes under the name of Lagrangian formalism of LGTs \cite{Creutz1985,Rothe}. Alternatively, as considered in the main text, the Hamiltonian formalism can be adopted \cite{Kogut-Susskind}, where spatial dimensions are discretized while time remains continuous. In this formulation, the states of the theory are constrained to satisfy the Gauss law.

%%%%%%%%%%%%%%%%%%%%%%%%%%%%%%%%%%%%%%%%%%%%%%%%%%%%%%%%%%%%%%%%%%%%%%%%%%%%

\section{\label{appendix_CT_dualization}Dualization of the single periodic plaquette}
In this Appendix we report the detailed derivation of the dualization and encoding of degrees of freedom for the case of the minimal torus, following the steps depicted in Fig. \ref{basic_CT + plaquette_changeofvars}. We use the canonical transformations (CTs) introduced in \cite{Mathur_PRD2023}, and build an iterative procedure to pass from links to loops and open strings that works in a square lattice with PBC, realizing a one-to-one mapping preserving the number of degrees of freedom. To make cleaner the procedure, we split the CTs in two groups: the first one is associated to the definition of independent plaquette operators, while the second to the large loops $\mathcal{L}_{\hat{x},\hat{y}}$ and the open strings.

%%%%%%%%%%%%%%%%%%%%%%%%%%%%%%%%%%%%%%%%%%%%%%%%%%%%%%%%%%%%%%%%%%%%%%%%%%%%

\subsection{First set of transformations: plaquette loop operators}
After the first set of CTs, we are left with the strings in the central panel of Fig. \ref{basic_CT + plaquette_changeofvars}(b). In terms of the original link variables, we have
\begin{equation}
    W_{\bm{A}}=U_{\hat{x}}(\bm{A})U_{\hat{y}}(\bm{B})U^\dagger_{\hat{x}}(\bm{D})U^\dagger_{\hat{y}}(\bm{A}),
    \label{W(A)_def}
\end{equation}
\begin{equation}
    W_{\bm{B}}=U^\dagger_{\hat{y}}(\bm{C})U_{\hat{x}}(\bm{C})U_{\hat{y}}(\bm{D})U^\dagger_{\hat{x}}(\bm{B}),
    \label{W(B)_def}
\end{equation}
\begin{equation}
    W_{\bm{C}}=U^\dagger_{\hat{y}}(\bm{B})U_{\hat{x}}(\bm{B})U_{\hat{y}}(\bm{A})U^\dagger_{\hat{x}}(\bm{C}),
    \label{W(C)_def}
\end{equation}
with conjugate electric fields
\begin{equation}
    \mathcal{E}_L(\bm{A})=E_{L,\hat{x}}(\bm{A}),\qquad \mathcal{E}_R(\bm{A})=-W^\dagger_{\bm{A}}\mathcal{E}_L(\bm{A})W_{\bm{A}},
    \label{E(A)_plaq_def}
\end{equation}
\begin{equation}
    \mathcal{E}_L(\bm{B})=E_{R,\hat{y}}(\bm{B}),\qquad \mathcal{E}_R(\bm{B})=-W^\dagger_{\bm{B}}\mathcal{E}_L(\bm{B})W_{\bm{B}},
    \label{E(B)_plaq_def}
\end{equation}
\begin{equation}
    \mathcal{E}_L(\bm{C})=E_{R,\hat{y}}(\bm{C})-U^\dagger_{\hat{y}}(\bm{B})E_{R,\hat{x}}(\bm{B})U_{\hat{y}}(\bm{B}),\qquad \mathcal{E}_R(\bm{C})=-W^\dagger_{\bm{C}}\mathcal{E}_L(\bm{C})W_{\bm{C}}.
    \label{E(C)_plaq_def}
\end{equation}

We observe that we need to define the origin of the two plaquette loops $W_{\bm{B}}$, $W_{\bm{C}}$ in the top left corner, instead than in the bottom left one as happens for $W_{\bm{A}}$. This does not happen if open boundaries are considered: here, when four plaquettes are considered, there are four links more, associated to the physical boundaries of the lattice, and every plaquette loop can be closed using the same convention \cite{MathurPRD2015,Mathur_PLB2015,Mathur_PRD2023}. In the periodic case, to avoid inconsistencies and use the same basic CT, we keep the same orientation and adopt different origins. As a final comment, there are five remaining links, according to the central panel of Fig. \ref{basic_CT + plaquette_changeofvars}(b), with redefined canonical electric fields. As we are going to show, they are recombined to obtain the two large loops and three open strings.

%%%%%%%%%%%%%%%%%%%%%%%%%%%%%%%%%%%%%%%%%%%%%%%%%%%%%%%%%%%%%%%%%%%%%%%%%%%%

\subsection{Second set of transformations: large loop and open strings operators}
We keep using the transformation rules in Eqs. \eqref{basic_CT_links}, \eqref{basis_CT_electricfields} to merge the remaining links into large loops wrapping the minimal torus in both the spatial directions. The final result of the CTs is showed in the right panel Fig. \ref{basic_CT + plaquette_changeofvars}(b). The two loops are wrapping the lattice horizontally ($\mathcal{L}_{\hat{x}}(\bm{C})$) and vertically ($\mathcal{L}_{\hat{y}}(\bm{A})$), and are defined as
\begin{equation}
    \mathcal{L}_{\hat{x}}(\bm{C})=U^\dagger_{\hat{x}}(\bm{D})U^\dagger_{\hat{x}}(\bm{C}),\qquad\qquad\mathcal{L}_{\hat{y}}(\bm{A})=U^\dagger_{\hat{y}}(\bm{D})U^\dagger_{\hat{y}}(\bm{A}).
\end{equation}
The canonical electric fields are \footnote{From now on we omit the expression of the right electric field, as they can be computed by means of Eq. \eqref{relation_left_right_electricfields}.}
\begin{equation}
    (E_{\mathcal{L}})_{L,\hat{x}}(\bm{C})=_{R,\hat{x}}(\bm{C})-U^\dagger_{\hat{y}}(\bm{B})U^\dagger_{\hat{x}}(\bm{A})\mathcal{E}_L(\bm{A})U_{\hat{x}}(\bm{A})U_{\hat{y}}(\bm{B}),
\end{equation}
\begin{equation}
    (E_{\mathcal{L}})_{L,\hat{y}}(\bm{A})=E_{R,\hat{y}}(\bm{A})-U^\dagger_{\hat{x}}(\bm{B})\mathcal{E}_R(\bm{B})U_{\hat{x}}(\bm{B}),
\end{equation}
The remaining three open strings are defined as
\begin{equation}
    \mathcal{T}(\bm{B})=U_{\hat{x}}(\bm{B})U_{\hat{y}}(\bm{C}),\qquad \bar{U}_{\hat{x}}(\bm{C})=U_{\hat{x}}(\bm{C}),\qquad \bar{U}_{\hat{y}}(\bm{A})=U_{\hat{y}}(\bm{A}),
\end{equation}
with canonical electric fields
\begin{equation}
    (E_{\tau})_{L,\hat{x}}(\bm{B})=E_{L,\hat{x}}(\bm{B})+E_{L,\hat{y}}(\bm{B})+E_{R,\hat{x}}(\bm{B})-\mathcal{E}_R(\bm{B}),
\end{equation}
\begin{align}
    \bar{E}_{L,\hat{x}}(\bm{C})=&E_{L,\hat{y}}(\bm{C})+E_{L,\hat{x}}(\bm{C})-\mathcal{L}^\dagger_{\hat{x}}(\bm{C})U^\dagger_{\hat{y}}(\bm{B})U^\dagger_{\hat{x}}(\bm{A})\mathcal{E}_L(\bm{A})U_{\hat{x}}(\bm{A})U_{\hat{y}}(\bm{B})\mathcal{L}_{\hat{x}}(\bm{C})\\
    \nonumber
    &-\mathcal{E}_R(\bm{C})+\mathcal{L}^\dagger_{\hat{x}}(\bm{C})E_{R,\hat{x}}(\bm{C})\mathcal{L}_{\hat{x}}(\bm{C}),
\end{align}
\begin{align}
    \bar{E}_{L,\hat{y}}(\bm{A})=&-U^\dagger_{\hat{x}}(\bm{B})(E_{\tau})_{L,\hat{x}}(\bm{B})U_{\hat{x}}(\bm{B})+U_{\hat{y}}(\bm{A})U^\dagger_{\hat{x}}(\bm{C})U_{\hat{y}}(\bm{C})\mathcal{E}_L(\bm{B})U^\dagger_{\hat{y}}(\bm{C})U_{\hat{x}}(\bm{C})U^\dagger_{\hat{y}}(\bm{A})\\
    \nonumber
    &-U_{\hat{y}}(\bm{A})E_{L,\hat{y}}(\bm{D})U^\dagger_{\hat{y}}(\bm{A})+E_{L,\hat{y}}(\bm{A})-\mathcal{E}_R(\bm{A})-U^\dagger_{\hat{x}}(\bm{B})U_{\hat{y}}(\bm{B})\mathcal{E}_L(\bm{C})U^\dagger_{\hat{y}}(\bm{B})U_{\hat{x}}(\bm{B}).
\end{align}
In the obtained expressions we recognize the initial Kogut--Susskind electric fields
\begin{equation}
    \mathcal{I}\equiv\{E_{L,\hat{x}}(\bm{A}),E_{R,\hat{y}}(\bm{A}),E_{L,\hat{y}}(\bm{A}),E_{R,\hat{y}}(\bm{B}),E_{L,\hat{x}}(\bm{B}),E_{R,\hat{y}}(\bm{C}),E_{R,\hat{x}}(\bm{C}),E_{L,\hat{x}}(\bm{C})\},
    \label{initial_set}
\end{equation}
and all the others can be expressed by means of their parallel transports \footnote{Their expression, which is only needed in intermediate steps of computations appearing in the next Sections, is reported here for completeness:
\begin{equation}
    E_{R,\hat{x}}(\bm{B})=-U^\dagger_{\hat{x}}(\bm{A})E_{L,\hat{x}}(\bm{A})U_{\hat{x}}(\bm{A})=-U^\dagger_{\hat{x}}(\bm{A})\mathcal{E}_L(\bm{A})U_{\hat{x}}(\bm{A}),
\end{equation}
\begin{equation}
    E_{L,\hat{y}}(\bm{C})=-U_{\hat{y}}(\bm{C})E_{R,\hat{y}}(\bm{B})U^\dagger_{\hat{y}}(\bm{C})=-U_{\hat{y}}(\bm{C})\mathcal{E}_L(\bm{B})U^\dagger_{\hat{y}}(\bm{C}),
\end{equation}
\begin{equation}
    E_{L,\hat{y}}(\bm{B})=-U_{\hat{y}}(\bm{B})\mathcal{E}_L(\bm{C})U^\dagger_{\hat{y}}(\bm{B})+U^\dagger_{\hat{x}}(\bm{A})\mathcal{E}_L(\bm{A})U_{\hat{x}}(\bm{A}),
\end{equation}
\begin{equation}
    E_{L,\hat{y}}(\bm{D})=-U_{\hat{y}}(\bm{D})[(E_\mathcal{L})_{L,\hat{y}}(\bm{A})+U^\dagger_{\hat{x}}(\bm{B})\mathcal{E}_R(\bm{B})U_{\hat{x}}(\bm{B})]U^\dagger_{\hat{y}}(\bm{D}).
\end{equation}
}.
The one-to-one mapping of degrees of freedom is now complete:
\begin{equation}
    \mathcal{I}\longleftrightarrow\{\mathcal{E}_L(\bm{A}),(E_{\mathcal{L}})_{L,\hat{y}}(\bm{A}),\bar{E}_{L,\hat{y}}(\bm{A}),\mathcal{E}_L(\bm{B}),(E_{\tau})_{L,\hat{x}}(\bm{B}),\mathcal{E}_L(\bm{C}),(E_{\mathcal{L}})_{L,\hat{x}}(\bm{C}),\bar{E}_{L,\hat{x}}(\bm{C})\}.
\end{equation}
The choice of the open strings is done to \textit{not} have new independent electric fields in $\bm{D}$. For this reason, this can be thought as the \textit{reference point} in the minimal torus.

Moreover, from the above expressions of the electric fields we notice that the inversion relations are immediate, as any independent electric field in the initial and final sets of variables appears with unit coefficient in the formulas. The dual loop variables, obtained through CTs, satisfy the canonical commutation relations. Explicitly, this means to have
\begin{equation}
    [\mathcal{E}^a_L(\bm{n}),\mathcal{W}_{ij}(\bm{m})]=-\delta_{\bm{mn}}[T^a\mathcal{W}(\bm{n})]_{ij},
\end{equation}
\begin{equation}
    [\mathcal{E}^a_R(\bm{n}),\mathcal{W}_{ij}(\bm{m})]=\delta_{\bm{mn}}[\mathcal{W}(\bm{n})T^a]_{ij}
\end{equation}
for any $\mathcal{E}\in\mathcal{I}$ attached to its loop $\mathcal{W}(\bm{n})$. We have as well that
\begin{equation}
    [\mathcal{E}^a_{L,R}(\bm{n}),\mathcal{E}^b_{L,R}(\bm{m})]=if^{abc}\mathcal{E}^c_{L,R}(\bm{n})\delta_{\bm{mn}},
    \label{reformulated_Lie_algebra}
\end{equation}
that is, the loop and open string electric fields satisfy the $SU(N)$ Lie algebra.

%%%%%%%%%%%%%%%%%%%%%%%%%%%%%%%%%%%%%%%%%%%%%%%%%%%%%%%%%%%%%%%%%%%%%%%%%%%%
\subsection{\label{gausslaw}Explicit resolution of the Gauss law}
We solve the independent Gauss laws in favor of the loop electric fields, to get rid of the open strings ending in the reference point. Since the change of variables has been done using CTs, which are linear in the electric fields, the reformulated local constraints are the sums of the new electric fields attached to a given lattice point. In terms of the loop electric fields, they read
\begin{equation}
    \mathcal{G}({\bm{A}})\equiv\mathcal{E}_L(\bm{A})+\mathcal{E}_R(\bm{A})+(E_{\mathcal{L}})_{L,\hat{y}}(\bm{A})+(E_{\mathcal{L}})_{R,\hat{y}}(\bm{A})+\bar{E}_{L,\hat{y}}(\bm{A})=0,
    \label{gauss_A}
\end{equation}
\begin{equation}
    {\cal G}(\bm{B})\equiv\mathcal{E}_L(\bm{B})+\mathcal{E}_R(\bm{B})+(E_{\tau})_{L,\hat{x}}(\bm{B})=0,
    \label{gauss_B}
\end{equation}
\begin{equation}
    \mathcal{G}({\bm{C}})\equiv\mathcal{E}_L(\bm{C})+\mathcal{E}_R(\bm{C})+(E_{\mathcal{L}})_{L,\hat{x}}(\bm{C})+(E_{\mathcal{L}})_{R,\hat{x}}(\bm{C})+\bar{E}_{L,\hat{x}}(\bm{C})=0,
    \label{gauss_C}
\end{equation}
while the Gauss law at the reference point is
\begin{equation}
    \mathcal{G}(\bm{D})\equiv\bar{E}_{R,\hat{y}}(\bm{D})+(E_{\tau})_{R,\hat{y}}(\bm{D})+\bar{E}_{R,\hat{x}}(\bm{D})=0
    \label{gauss_D}
\end{equation}
and is not independent from the others, as its computation can be reconduced to the definition of $\bar{E}_{L,\hat{y}}(\bm{A})$. We refer to Appendix \ref{verify_newgauss} for all the computations and checks regarding the dualized local constraints.

We explicitly solve the three Gauss laws in Eq. \eqref{gauss_A}, \eqref{gauss_B} and \eqref{gauss_C} to eliminate the electric fields of the open strings, i.e.,
\begin{equation}
    \bar{E}_{L,\hat{y}}(\bm{A})=-[\mathcal{E}_L(\bm{A})+\mathcal{E}_R(\bm{A})+(E_{\mathcal{L}})_{L,\hat{y}}(\bm{A})+(E_{\mathcal{L}})_{R,\hat{y}}(\bm{A})],
    \label{gauss_A_solve}
\end{equation}
\begin{equation}
    (E_{\tau})_{L,\hat{x}}(\bm{B})=-[\mathcal{E}_L(\bm{B})+\mathcal{E}_R(\bm{B})],
    \label{gauss_B_solve}
\end{equation}
\begin{equation}
    \bar{E}_{L,\hat{x}}(\bm{C})=-[\mathcal{E}_L(\bm{C})+\mathcal{E}_R(\bm{C})+(E_{\mathcal{L}})_{L,\hat{x}}(\bm{C})+(E_{\mathcal{L}})_{R,\hat{x}}(\bm{C})].
    \label{gauss_C_solve}
\end{equation}
As a last step to obtain the physical variables, we use the gauge freedom to set the open strings to the identity in the gauge group. This allows us to write the Hamiltonian and the wave functions only as a functions of the loops. If we label with $\mathcal{W}$ the set of closed loops, and with $\mathcal{S}$ the set of open strings,
\begin{equation}
    \{\underbrace{W_{\bm{A}},\mathcal{L}_{\hat{y}}(\bm{A}),W_{\bm{B}},W_{\bm{C}},\mathcal{L}_{\hat{x}}(\bm{C})}_{\equiv\mathcal{W}},\underbrace{\bar{U}_{\hat{y}}(\bm{A}),\mathcal{T}(\bm{B}),\bar{U}_{\hat{x}}(\bm{C})}_{\equiv\mathcal{S}}\}=\mathcal{W}\oplus\mathcal{S},
    \label{singleplaq_W_S_defs}
\end{equation}
we act with proper gauge transformations to fix $\mathcal{T}=\mathbb{1}_{\mathcal{G}},\;\forall\;\mathcal{T}\in\mathcal{S}$. On the basis states of the system, which we generically label as
\begin{equation}
    |\psi\rangle=|\mathcal{W}_1,\ldots\mathcal{W}_n,\mathcal{T}_1,\ldots,\mathcal{T}_m\rangle,\qquad\mathcal{W}_i\in\mathcal{W},\;\mathcal{T}_i\in\mathcal{S},
\end{equation}
we use a product of gauge transformations $\Theta_{\mathcal{T}_i}(\bm{n})$ such that
\begin{equation}
    |\mathcal{W}_1,\ldots\mathcal{W}_n,\mathcal{T}_1,\ldots,\mathcal{T}_m\rangle\qquad\rightarrow\qquad|\psi'\rangle=|\mathcal{W}'_1,\ldots\mathcal{W}'_n,\mathbb{1}_{\mathcal{G}},\ldots,\mathbb{1}_{\mathcal{G}}\rangle,
\end{equation}
where $\mathcal{W}'_i$ are the corresponding gauge transformed loops. This procedure is general, and holds for any dimension of the sets $\mathcal{W},\;\mathcal{S}$ \cite{Bauer2023}. For the minimal torus, the above product of gauge transformations is
\begin{equation}
    \mathbb{G}=e^{i\theta^a\bar{E}^a_{L,\hat{y}}(\bm{A})}e^{i\theta^a (E^a_{\tau})_{L,\hat{x}}(\bm{B})}e^{i\theta^a\bar{E}^a_{L,\hat{x}}(\bm{C})}.
    \label{GT_fix_openstrings}
\end{equation}
This open strings gauge fixing is compatible with the procedure outlined by Creutz \cite{Creutz1977}, as we can define a disconnected maximal tree gauge made by all the open strings $\mathcal{T}\in\mathcal{S}$. As it can be verified from the right plot of Fig. \ref{basic_CT + plaquette_changeofvars}(b), the open strings form a maximal tree, i.e., a set of disconnected links such that no more links can be added without forming a closed loop \footnote{Another possibility to visualize this is to go back the initial link variables $U_\mu(\bm{n})$, and identify the maximal tree corresponding to our procedure.}. Therefore, all of them can be fixed to an arbitrary element of the gauge group, which we choose to be the identity.

%%%%%%%%%%%%%%%%%%%%%%%%%%%%%%%%%%%%%%%%%%%%%%%%%%%%%%%%%%%%%%%%%%%%%%%%%%%%

\subsection{Final form of the inverse relations}
We insert Eqs. \eqref{gauss_A_solve}, \eqref{gauss_B_solve} and \eqref{gauss_C_solve} into the expressions for the dual variables and invert the relations to isolate the set of electric fields identified in Eq. \eqref{initial_set}. Due to the explicit resolution of the gauge symmetry, we express the eight KS electric fields as a function of five loop electric fields. This dualization and encoding procedure is performed as well for the links, allowing us to express them solely in terms of the set $\mathcal{W}$ introduced in Eq. \eqref{singleplaq_W_S_defs} by taking into account the gauge fixing of the open strings. The obtained expressions are the Eqs. \eqref{inverse_links}, \eqref{inverse_E_1}, \eqref{inverse_E_2}, \eqref{inverse_E_3} and \eqref{inverse_E_4} reported in the main text.

%%%%%%%%%%%%%%%%%%%%%%%%%%%%%%%%%%%%%%%%%%%%%%%%%%%%%%%%%%%%%%%%%%%%%%%%%%%%

\section{\label{verify_newgauss}Verifying the dual Gauss laws}
In this Appendix we explicitly verify that the dual electric fields $\mathcal{E}\in\mathcal{I}$ satisfy Eqs. \eqref{gauss_A}, \eqref{gauss_B}, \eqref{gauss_C} and \eqref{gauss_D}.
\subsection{Gauss law in $A$}
We have that
\begin{equation}
    \begin{aligned}
        \mathcal{G}(\bm{A})&=E_{L,\hat{x}}(\bm{A})+\mathcal{E}_R(\bm{A})+E_{R,\hat{y}}(\bm{A})-U^\dagger_{\hat{x}}(\bm{B})\mathcal{E}_R(\bm{B})U_{\hat{x}}(\bm{B})+(E_{\mathcal{L}})_{R,\hat{y}}(\bm{A})\\
        &+E_{L,\hat{y}}(\bm{A})-U^\dagger_{\hat{x}}(\bm{B})[E_{L,\hat{x}}(\bm{B})+E_{L,\hat{y}}(\bm{B})+E_{R,\hat{x}}(\bm{B})-\mathcal{E}_R(\bm{B})]U_{\hat{x}}(\bm{B})\\
        &+U_{\hat{y}}(\bm{A})U^\dagger_{\hat{x}}(\bm{C})U_{\hat{y}}(\bm{C})\mathcal{E}_L(\bm{B})U^\dagger_{\hat{y}}(\bm{C})U_{\hat{x}}(\bm{C})U^\dagger_{\hat{y}}(\bm{A})-U_{\hat{y}}(\bm{A})E_{L,\hat{y}}(\bm{D})U^\dagger_{\hat{y}}(\bm{A})\\
        &-\mathcal{E}_R(\bm{A})-U^\dagger_{\hat{x}}(\bm{B})U_{\hat{y}}(\bm{B})\mathcal{E}_L(\bm{C})U^\dagger_{\hat{y}}(\bm{B})U_{\hat{x}}(\bm{B}).
    \end{aligned}
\end{equation}
Some of the terms cancel out directly, in particular we have that $-U^\dagger_{\hat{x}}(\bm{B})E_{L,\hat{x}}(\bm{B})U_{\hat{x}}(\bm{B})=E_{R,\hat{x}}(\bm{A})$ and all the four electric fields in $\bm{A}$ sum to zero, due to the Gauss law for the KS electric fields. By using $E_{L,\hat{y}}(\bm{D})=-U_{\hat{y}}(\bm{D})[(E_\mathcal{L})_{L,\hat{y}}(\bm{A})+U^\dagger_{\hat{x}}(\bm{B})\mathcal{E}_R(\bm{B})U_{\hat{x}}(\bm{B})]U^\dagger_{\hat{y}}(\bm{D})$ we can write
\begin{equation}
    \begin{aligned}
        \mathcal{G}(\bm{A})&=(E_{\mathcal{L}})_{R,\hat{y}}(\bm{A})-U^\dagger_{\hat{x}}(\bm{B})[E_{L,\hat{y}}(\bm{B})+E_{R,\hat{x}}(\bm{B})]U_{\hat{x}}(\bm{B})\\
        &+U_{\hat{y}}(\bm{A})U_{\hat{y}}(\bm{D})[(E_\mathcal{L})_{L,\hat{y}}(\bm{A})+U^\dagger_{\hat{x}}(\bm{B})\mathcal{E}_R(\bm{B})U_{\hat{x}}(\bm{B})]U^\dagger_{\hat{y}}(\bm{D})U^\dagger_{\hat{y}}(\bm{A})\\
        &+U_{\hat{y}}(\bm{A})U^\dagger_{\hat{x}}(\bm{C})U_{\hat{y}}(\bm{C})\mathcal{E}_L(\bm{B})U^\dagger_{\hat{y}}(\bm{C})U_{\hat{x}}(\bm{C})U^\dagger_{\hat{y}}(\bm{A})-U^\dagger_{\hat{x}}(\bm{B})U_{\hat{y}}(\bm{B})\mathcal{E}_L(\bm{C})U^\dagger_{\hat{y}}(\bm{B})U_{\hat{x}}(\bm{B}).
    \end{aligned}
\end{equation}
We make use of $(E_{\mathcal{L}})_{R,\hat{y}}(\bm{A})=-U_{\hat{y}}(\bm{A})U_{\hat{y}}(\bm{D})(E_\mathcal{L})_{L,\hat{y}}(\bm{A})
U^\dagger_{\hat{y}}(\bm{D})U^\dagger_{\hat{y}}(\bm{A})$ to simplify these two terms. Moreover, from the definition $\mathcal{E}_R(\bm{B})=-W^\dagger_{\bm{B}}\mathcal{E}_L(\bm{B})W_{\bm{B}}$, we observe that the path
\begin{equation}
    W_{\bm{B}}U_{\hat{x}}(\bm{B})U^\dagger_{\hat{y}}(\bm{D})U^\dagger_{\hat{y}}(\bm{A})=U^\dagger_{\hat{y}}(\bm{C})U_{\hat{x}}(\bm{C})U^\dagger_{\hat{y}}(\bm{A})
\end{equation}
and we are left with
\begin{equation}
    \mathcal{G}(\bm{A})=-U^\dagger_{\hat{x}}(\bm{B})[E_{L,\hat{y}}(\bm{B})+E_{R,\hat{x}}(\bm{B})]U_{\hat{x}}(\bm{B})-U^\dagger_{\hat{x}}(\bm{B})U_{\hat{y}}(\bm{B})\mathcal{E}_L(\bm{C})U^\dagger_{\hat{y}}(\bm{B})U_{\hat{x}}(\bm{B}),
\end{equation}
which is identically zero, by using $E_{R,\hat{x}}(\bm{B})=-U^\dagger_{\hat{x}}(\bm{A})\mathcal{E}_L(\bm{A})U_{\hat{x}}(\bm{A})$ and $E_{L,\hat{y}}(\bm{B})=-U_{\hat{y}}(\bm{B})\mathcal{E}_L(\bm{C})U^\dagger_{\hat{y}}(\bm{B})+U^\dagger_{\hat{x}}(\bm{A})\mathcal{E}_L(\bm{A})U_{\hat{x}}(\bm{A})$.

\subsection{Gauss law in $B$}
This is the easiest case, since it already follows from the definitions of the dual electric fields that
\begin{equation}
    \mathcal{G}(\bm{B})=E_{R,\hat{y}}(\bm{B})+E_{L,\hat{x}}(\bm{B})+E_{L,\hat{y}}(\bm{B})+E_{R,\hat{x}}(\bm{B})=0.
\end{equation}

\subsection{Gauss law in $C$}
By using the definitions of loop electric fields we have
\begin{equation}
    \begin{aligned}
        \mathcal{G}(\bm{C})&=E_{R,\hat{y}}(\bm{C})-U^\dagger_{\hat{y}}(\bm{B})E_{R,\hat{x}}(\bm{B})U_{\hat{y}}(\bm{B})+\mathcal{E}_R(\bm{C})-U^\dagger_{\hat{y}}(\bm{B})U^\dagger_{\hat{x}}(\bm{A})\mathcal{E}_L(\bm{A})U_{\hat{x}}(\bm{A})U_{\hat{y}}(\bm{B})\\
        &+E_{R,\hat{x}}(\bm{C})+(E_{\mathcal{L}})_{R,\hat{x}}(\bm{C})+E_{L,\hat{y}}(\bm{C})+E_{L,\hat{x}}(\bm{C})-\mathcal{E}_R(\bm{C})\\
        &+\mathcal{L}^\dagger_{\hat{x}}(\bm{C})E_{R,\hat{x}}(\bm{C})\mathcal{L}_{\hat{x}}(\bm{C})-\mathcal{L}^\dagger_{\hat{x}}(\bm{C})U^\dagger_{\hat{y}}(\bm{B})U^\dagger_{\hat{x}}(\bm{A})\mathcal{E}_L(\bm{A})U_{\hat{x}}(\bm{A})U_{\hat{y}}(\bm{B})\mathcal{L}_{\hat{x}}(\bm{C}).
    \end{aligned}
\end{equation}
Since $E_{R,\hat{x}}(\bm{B})=-U^\dagger_{\hat{x}}(\bm{A})\mathcal{E}_L(\bm{A})U_{\hat{x}}(\bm{A})$, we observe that the second and fourth terms of the right-hand side in the first line cancel out, as well as the two equal and opposite $\mathcal{E}_R(\bm{C})$. Moreover, we recognize that 
\begin{equation}
    (E_{\mathcal{L}})_{R,\hat{x}}(\bm{C})=-\mathcal{L}^\dagger_{\hat{x}}(\bm{C})E_{R,\hat{x}}(\bm{C})\mathcal{L}_{\hat{x}}(\bm{C})+\mathcal{L}^\dagger_{\hat{x}}(\bm{C})U^\dagger_{\hat{y}}(\bm{B})U^\dagger_{\hat{x}}(\bm{A})\mathcal{E}_L(\bm{A})U_{\hat{x}}(\bm{A})U_{\hat{y}}(\bm{B})\mathcal{L}_{\hat{x}}(\bm{C}),
\end{equation}
simplifying then with the entire last line. We are left finally with the Gauss law in terms of the initial variables, i.e.
\begin{equation}
    \mathcal{G}(\bm{C})=E_{R,\hat{y}}(\bm{C})+E_{R,\hat{x}}(\bm{C})+E_{L,\hat{y}}(\bm{C})+E_{L,\hat{x}}(\bm{C})=0.
\end{equation}

\subsection{Gauss law in $D$}
This computation depends on the other Gauss laws and loop electric fields definitions. Indeed we have
\begin{equation}
    \mathcal{G}(\bm{D})=-U^\dagger_{\hat{y}}(\bm{A})\bar{E}_{L,\hat{y}}(\bm{A})U_{\hat{y}}(\bm{A})-U^\dagger_{\hat{x}}(\bm{C})\bar{E}_{L,\hat{x}}(\bm{C})U_{\hat{x}}(\bm{C})-U^\dagger_{\hat{y}}(\bm{A})U^\dagger_{\hat{x}}(\bm{B})(E_\tau)_{L,\hat{x}}(\bm{B})U_{\hat{x}}(\bm{B})U_{\hat{y}}(\bm{A}),
\end{equation}
and multiplying on the left by $U_{\hat{y}}(\bm{A})$ and on the right by $U^\dagger_{\hat{y}}(\bm{A})$ we reduce to the computation of $\bar{E}_{L,\hat{y}}(\bm{A})$. What we have to check is that
\begin{equation}
    \bar{E}_{L,\hat{y}}(\bm{A})=-U^\dagger_{\hat{x}}(\bm{B})(E_\tau)_{L,\hat{x}}(\bm{B})U_{\hat{x}}(\bm{B})-U_{\hat{y}}(\bm{A})U^\dagger_{\hat{x}}(\bm{C})\bar{E}_{L,\hat{x}}(\bm{C})U_{\hat{x}}(\bm{C})U^\dagger_{\hat{y}}(\bm{A}).
    \label{check_barA_definition}
\end{equation}
Since the first term in the right-hand side is already present in the definition of $\bar{E}_{L,\hat{y}}(\bm{A})$, we expand the second term by using the definition of $\bar{E}_{L,\hat{x}}(\bm{C})$. We get
\begin{equation}
    \begin{aligned}
        U_{\hat{y}}(\bm{A})U^\dagger_{\hat{x}}(\bm{C})\bar{E}_{L,\hat{x}}(\bm{C})U_{\hat{x}}(\bm{C})U^\dagger_{\hat{y}}(\bm{A})=&-U_{\hat{y}}(\bm{A})U^\dagger_{\hat{x}}(\bm{C})[E_{L,\hat{y}}(\bm{C})+E_{L,\hat{x}}(\bm{C})\\
        &-\mathcal{L}^\dagger_{\hat{x}}(\bm{C})U^\dagger_{\hat{y}}(\bm{B})U^\dagger_{\hat{x}}(\bm{A})\mathcal{E}_L(\bm{A})U_{\hat{x}}(\bm{A})U_{\hat{y}}(\bm{B})\mathcal{L}_{\hat{x}}(\bm{C})\\
        &-\mathcal{E}_R(\bm{C})+\mathcal{L}^\dagger_{\hat{x}}(\bm{C})E_{R,\hat{x}}(\bm{C})\mathcal{L}_{\hat{x}}(\bm{C})]U_{\hat{x}}(\bm{C})U^\dagger_{\hat{y}}(\bm{A}),
    \end{aligned}
\end{equation}
which has five terms to be computed. Separately, for each of them, we have
\begin{equation}
    -U_{\hat{y}}(\bm{A})U^\dagger_{\hat{x}}(\bm{C})E_{L,\hat{y}}(\bm{C})U_{\hat{x}}(\bm{C})U^\dagger_{\hat{y}}(\bm{A})=U_{\hat{y}}(\bm{A})U^\dagger_{\hat{x}}(\bm{C})U_{\hat{y}}(\bm{C})\mathcal{E}_L(\bm{B})U^\dagger_{\hat{y}}(\bm{C})U_{\hat{x}}(\bm{C})U^\dagger_{\hat{y}}(\bm{A}),
    \label{(1)}
\end{equation}
\begin{equation}
    -U_{\hat{y}}(\bm{A})U^\dagger_{\hat{x}}(\bm{C})E_{L,\hat{x}}(\bm{C})U_{\hat{x}}(\bm{C})U^\dagger_{\hat{y}}(\bm{A})=U_{\hat{y}}(\bm{A})E_{R,\hat{x}}(\bm{D})U^\dagger_{\hat{y}}(\bm{A}),
    \label{(2)}
\end{equation}
\begin{equation}
    U_{\hat{y}}(\bm{A})U^\dagger_{\hat{x}}(\bm{C})\mathcal{L}^\dagger_{\hat{x}}(\bm{C})U^\dagger_{\hat{y}}(\bm{B})U^\dagger_{\hat{x}}(\bm{A})\mathcal{E}_L(\bm{A})U_{\hat{x}}(\bm{A})U_{\hat{y}}(\bm{B})\mathcal{L}_{\hat{x}}(\bm{C})U_{\hat{x}}(\bm{C})U^\dagger_{\hat{y}}(\bm{A})=-\mathcal{E}_R(\bm{A}),
    \label{(3)}
\end{equation}
\begin{equation}
    U_{\hat{y}}(\bm{A})U^\dagger_{\hat{x}}(\bm{C})\mathcal{E}_R(\bm{C})U_{\hat{x}}(\bm{C})U^\dagger_{\hat{y}}(\bm{A})=-U^\dagger_{\hat{x}}(\bm{B})U_{\hat{y}}(\bm{B})\mathcal{E}_L(\bm{C})U^\dagger_{\hat{y}}(\bm{B})U_{\hat{x}}(\bm{B}),
    \label{(4)}
\end{equation}
\begin{equation}
    -U_{\hat{y}}(\bm{A})U^\dagger_{\hat{x}}(\bm{C})\mathcal{L}^\dagger_{\hat{x}}(\bm{C})E_{R,\hat{x}}(\bm{C})\mathcal{L}_{\hat{x}}(\bm{C})U_{\hat{x}}(\bm{C})U^\dagger_{\hat{y}}(\bm{A})=U_{\hat{y}}(\bm{A})E_{L,\hat{x}}(\bm{D})U^\dagger_{\hat{y}}(\bm{A}),
    \label{(5)}
\end{equation}
where we simplified the paths $U_{\hat{y}}(\bm{A})U^\dagger_{\hat{x}}(\bm{C})\mathcal{L}^\dagger_{\hat{x}}(\bm{C})U^\dagger_{\hat{y}}(\bm{B})U^\dagger_{\hat{x}}(\bm{A})=W^\dagger_{\bm{A}}$, $W_{\bm{C}}U_{\hat{x}}(\bm{C})U^\dagger_{\hat{y}}(\bm{A})=U^\dagger_{\hat{y}}(\bm{B})U_{\hat{x}}(\bm{B})$ and $\mathcal{L}_{\hat{x}}(\bm{C})U_{\hat{x}}(\bm{C})U^\dagger_{\hat{y}}(\bm{A})=U^\dagger_{\hat{x}}(\bm{D})U^\dagger_{\hat{y}}(\bm{A})$.

By summing up Eqs. \eqref{(2)} and \eqref{(5)}, and using in the intermediate step the Gauss law in $\bm{D}$ for the KS electric fields, we have
\begin{equation}
    \begin{aligned}
        U_{\hat{y}}(\bm{A})[E_{L,\hat{x}}(\bm{D})+E_{R,\hat{x}}(\bm{D})]U^\dagger_{\hat{y}}(\bm{A})=&-U_{\hat{y}}(\bm{A})[E_{L,\hat{y}}(\bm{D})+E_{R,\hat{y}}(\bm{D})]U^\dagger_{\hat{y}}(\bm{A})\\
        &=E_{L,\hat{y}}(\bm{A})-U_{\hat{y}}(\bm{A})E_{L,\hat{y}}(\bm{D})U^\dagger_{\hat{y}}(\bm{A}).
    \end{aligned} 
\end{equation}
Finally, if we sum Eqs. \eqref{(1)}-\eqref{(5)} with the first term in the right-hand side of Eq. \eqref{check_barA_definition}, we get exactly the definition of $\bar{E}_{L,\hat{y}}(\bm{A})$, meaning that the Gauss law $\mathcal{G}(\bm{D})$ holds in terms of the dual variables.

\section{\label{electric_loop_commutators}Symmetries of the dual Hamiltonian}
We compute the commutator of the dual Hamiltonian with the loop electric fields $(E_{\mathcal{L}})_{L,\hat{y}}(\bm{A})$ and $(E_{\mathcal{L}})_{L,\hat{x}}(\bm{C})$, to show explicitly that the strings $\mathcal{L}_{\hat{y}}(\bm{A})$, $\mathcal{L}_{\hat{x}}(\bm{C})$ can not be considered as constants of motion.

We perform all the steps only for $\mathcal{L}_{\hat{y}}(\bm{A})$, and write the other one directly, since it is almost identical. First of all we observe that $[H,(E_{\mathcal{L}})_{L,\hat{y}}(\bm{A})]=[H_E,(E_{\mathcal{L}})_{L,\hat{y}}(\bm{A})]$, as the magnetic Hamiltonian does not contain the large loops. We consider separately the local and non-local electric contributions, respectively. For the first one we have
\begin{equation}
    [H_{E,\text{loc}},(E_{\mathcal{L}})_{L,\hat{y}}(\bm{A})]=g^2[(E_{\mathcal{L}})^2_{L,\hat{y}}(\bm{A}),(E_{\mathcal{L}})_{L,\hat{y}}(\bm{A})],
\end{equation}
and using the property $[A^2,B]=\{A,[A,B]\}$ together with the group algebra in Eq. \eqref{reformulated_Lie_algebra} we get
\begin{equation}
    [H_{E,\text{loc}},(E_{\mathcal{L}}^b)_{L,\hat{y}}(\bm{A})]=ig^2f^{abc}\{(E_{\mathcal{L}}^a)_{L,\hat{y}}(\bm{A}),(E_{\mathcal{L}}^c)_{L,\hat{y}}(\bm{A})\},
\end{equation}
where we explicitly wrote down the internal indices $a,\;b,\;c$ of the $SU(N)$ gauge group. We immediately observe that this contribution is zero, since the structure constants are completely antisymmetric in the group indices, and they are contracted with a commutator, which is a symmetric object. We therefore conclude that $[H_{E,\text{loc}},(E_{\mathcal{L}}^b)_{L,\hat{y}}(\bm{A})]=0$.

Regarding the commutator with the non-local electric Hamiltonian, things get more complicated due to the presence of parallel transporters. The only surviving terms are
\begin{align}
    \nonumber
    [H_{E,\text{non-loc}},(E_{\mathcal{L}})_{L,\hat{y}}(\bm{A})]=g^2&[\mathcal{F}_A[\mathcal{E}_{L,R}](E_{\mathcal{L}})_{L,\hat{y}}(\bm{A})+\mathcal{E}_L(\bm{C})\mathcal{R}[\mathcal{L}_{\hat{y}}(\bm{A})]\mathcal{E}_R(\bm{B})\\
    &+(E_{\mathcal{L}})_{L,\hat{x}}(\bm{C})\mathcal{R}[\mathcal{L}_{\hat{y}}(\bm{A})]\mathcal{E}_R(\bm{B}),(E_{\mathcal{L}})_{L,\hat{y}}(\bm{A})],
    \label{nonlocal_commutator_A}
\end{align}
where $\mathcal{F}_A[\mathcal{E}_{L,R}]\equiv\mathcal{E}_L(\bm{A})+\mathcal{E}_R(\bm{C})+\mathcal{E}_L(\bm{B})+\mathcal{E}_R(\bm{B})$. The terms not involving the parallel transport simplifies through Eq. \eqref{reformulated_Lie_algebra}, and do not vanish since $\mathcal{F}[\mathcal{E}_{L,R}]\neq0$. The terms with the parallel transport can be reduced to the computation of $[\mathcal{R}^{ab}[\mathcal{L}_{\hat{y}}(\bm{A})],(E_{\mathcal{L}})^c_{L,\hat{y}}(\bm{A})]$, which is non-vanishing in general if the group indices are not contracted \cite{Bauer2023}.  

The non-local electric Hamiltonian is the only contribution to the full commutator, therefore we conclude that
\begin{equation}
    [H,(E_{\mathcal{L}})_{L,\hat{y}}(\bm{A})]\neq 0,
    \label{vertical_loop_H_commutator}
\end{equation}
meaning that the vertical loop $\mathcal{L}_{\hat{y}}(\bm{A})$ is not a constant of motion.

The computation for the horizontal loop $\mathcal{L}_{\hat{x}}(\bm{C})$ can be done on the same lines. The magnetic and local electric Hamiltonians trivially commute with the conjugate electric field $(E_{\mathcal{L}})^c_{L,\hat{x}}(\bm{C})$. The non-trivial contribution comes from the non-local electric part, which is
\begin{equation}
    [H_{E,\text{non-loc}},(E_{\mathcal{L}})_{L,\hat{x}}(\bm{C})]=g^2[\mathcal{F}_C[\mathcal{E}_{L,R}](E_{\mathcal{L}})_{L,\hat{x}}(\bm{C})+\mathcal{E}_L(\bm{C})\mathcal{R}[\mathcal{L}^\dagger_{\hat{x}}(\bm{C})]\mathcal{E}_R(\bm{A}),(E_{\mathcal{L}})_{L,\hat{x}}(\bm{C})],
    \label{nonlocal_commutator_C}
\end{equation}
where $\mathcal{F}_C[\mathcal{E}_{L,R}]\equiv\mathcal{E}_L(\bm{C})+\mathcal{R}[\mathcal{L}_{\hat{y}}(\bm{A})]\mathcal{E}_R(\bm{B})$. The same considerations of the vertical large loop apply here, and we conclude that 
\begin{equation}
    [H,(E_{\mathcal{L}})_{L,\hat{x}}(\bm{C})]\neq 0.
    \label{horizontal_loop_H_commutator}
\end{equation}

%%%%%%%%%%%%%%%%%%%%%%%%%%%%%%%%%%%%%%%%%%%%%%%%%%%%%%%%%%%%%%%%%%%%%%%%%%%%

\section{\label{generalization_square_lattice}Generalization to arbitrary periodic square lattices}
We present the extension of the reformulation to an arbitrary square lattice of size $N_x\times N_y$. Since the mathematical expressions of the loop electric fields get more involved, we omit them and refer only to the graphical construction of the loop variables. However, the CTs are essentially iterations of Eqs. \eqref{basic_CT_links}, \eqref{basis_CT_electricfields}, and the dual electric fields can be reconstructed from the sequence of CTs by summing and parallel transporting accordingly the link electric fields. 

To understand the general structure of CTs for a general torus, we show how to extend the single periodic plaquette along the vertical and horizontal directions. The combination of these two operational blocks leads to the dualization for the $N_x\times N_y$ periodic lattice.

\subsection{\label{vertical_plaquettes}Vertical extension}
\begin{figure}[h]
    \centering
    \includegraphics[width=0.45\linewidth]{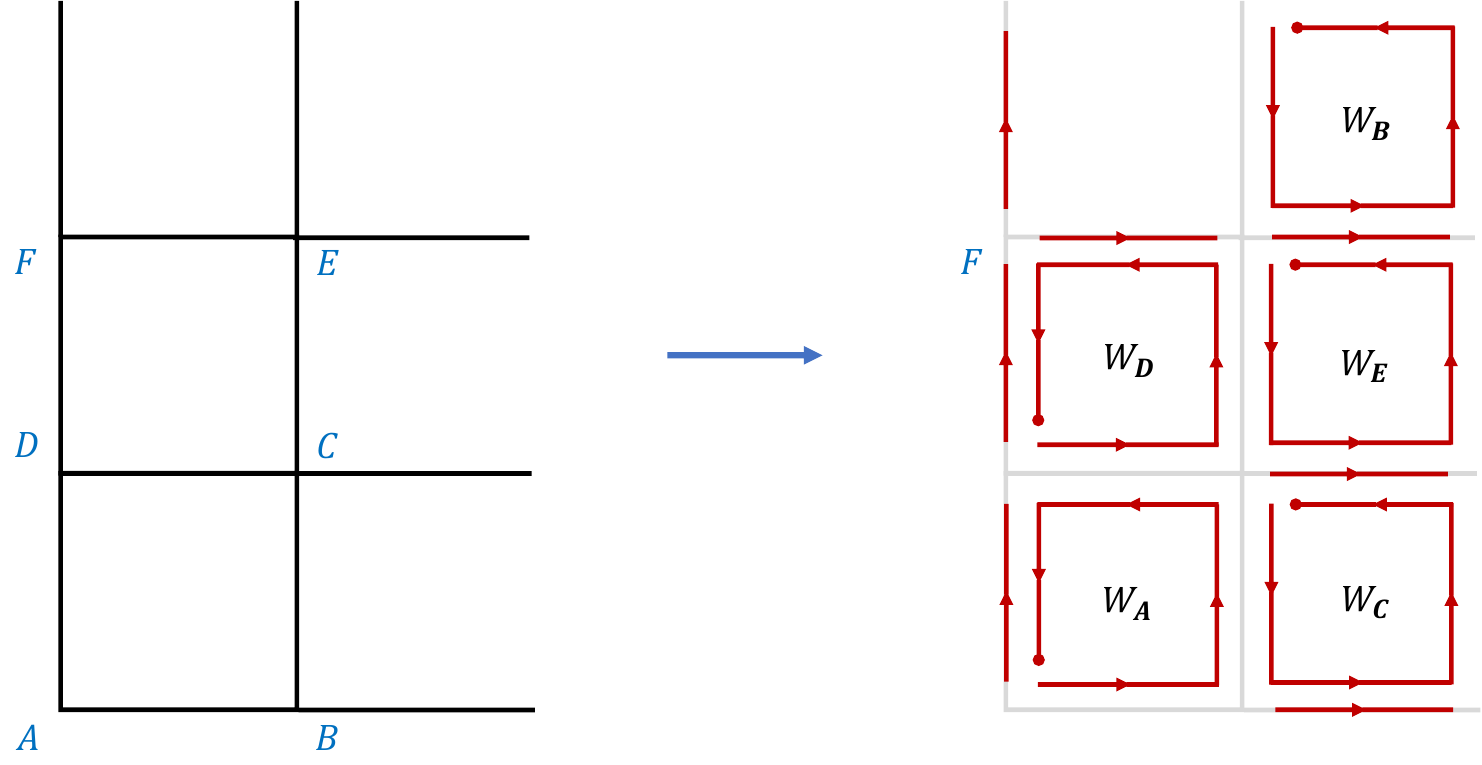}
    \qquad\qquad
    \includegraphics[width=0.45\linewidth]{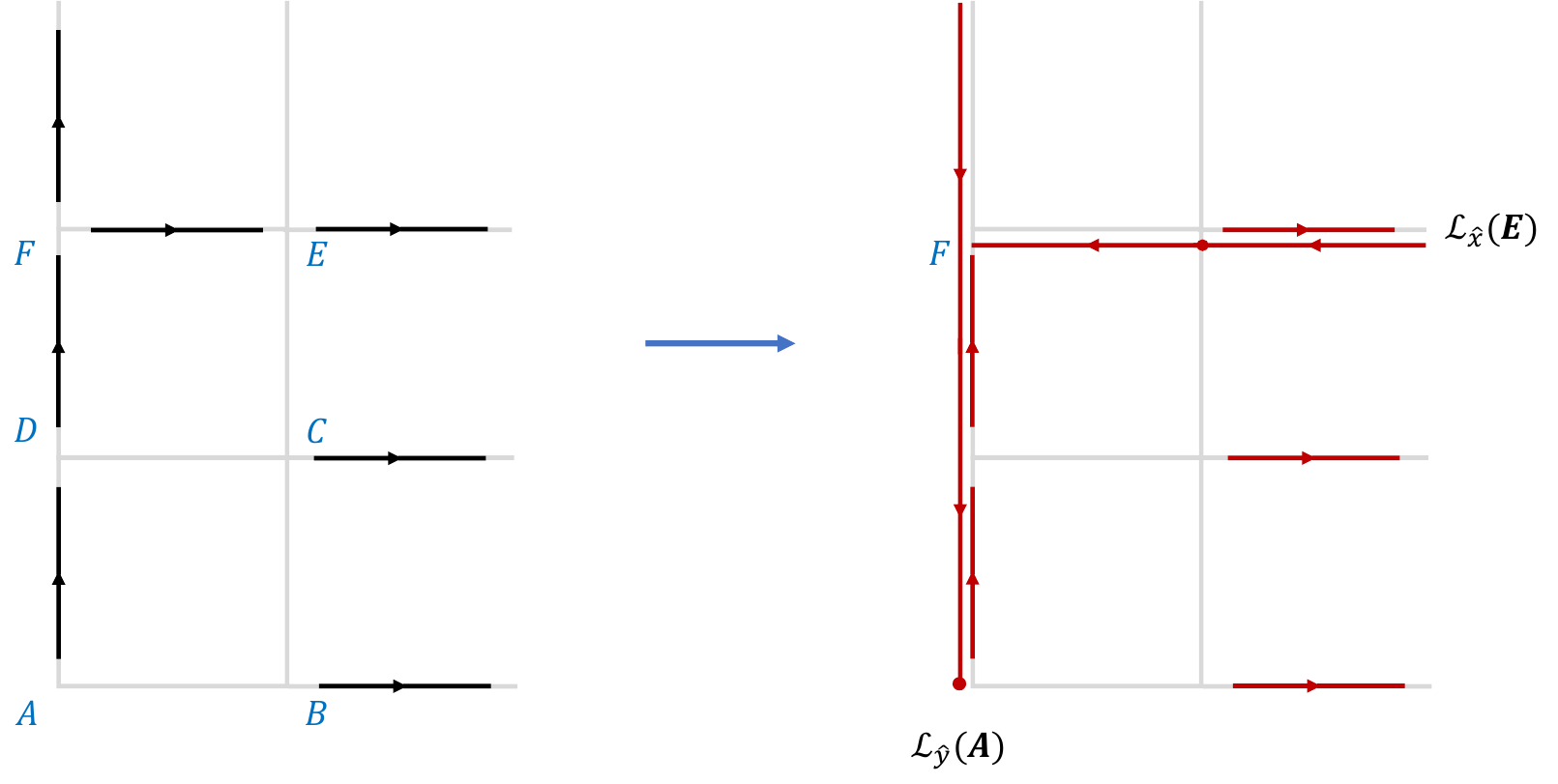}
    \caption{Graphical representation of canonical transformations to obtain five independent plaquettes $W_{\bm{n}}$ (left plot) and two large loops $\mathcal{L}_{\hat{x},\hat{y}}$ (right plot). In the final lattices, we report only the reference site $\bm{F}$.}
    \label{vertical_extension_CTs_1}
\end{figure}

We firstly count the degrees of freedom: initially, we have 12 links and 5 independent Gauss laws (Fig. \ref{vertical_extension_CTs_1}, left panel). After the CTs, we have 5 plaquettes, 2 large loops wrapping the lattice and a set of 5 open strings. All of the final degrees of freedom are physical and independent.
\begin{figure}[h]
    \centering
    \includegraphics[width=0.45\linewidth]{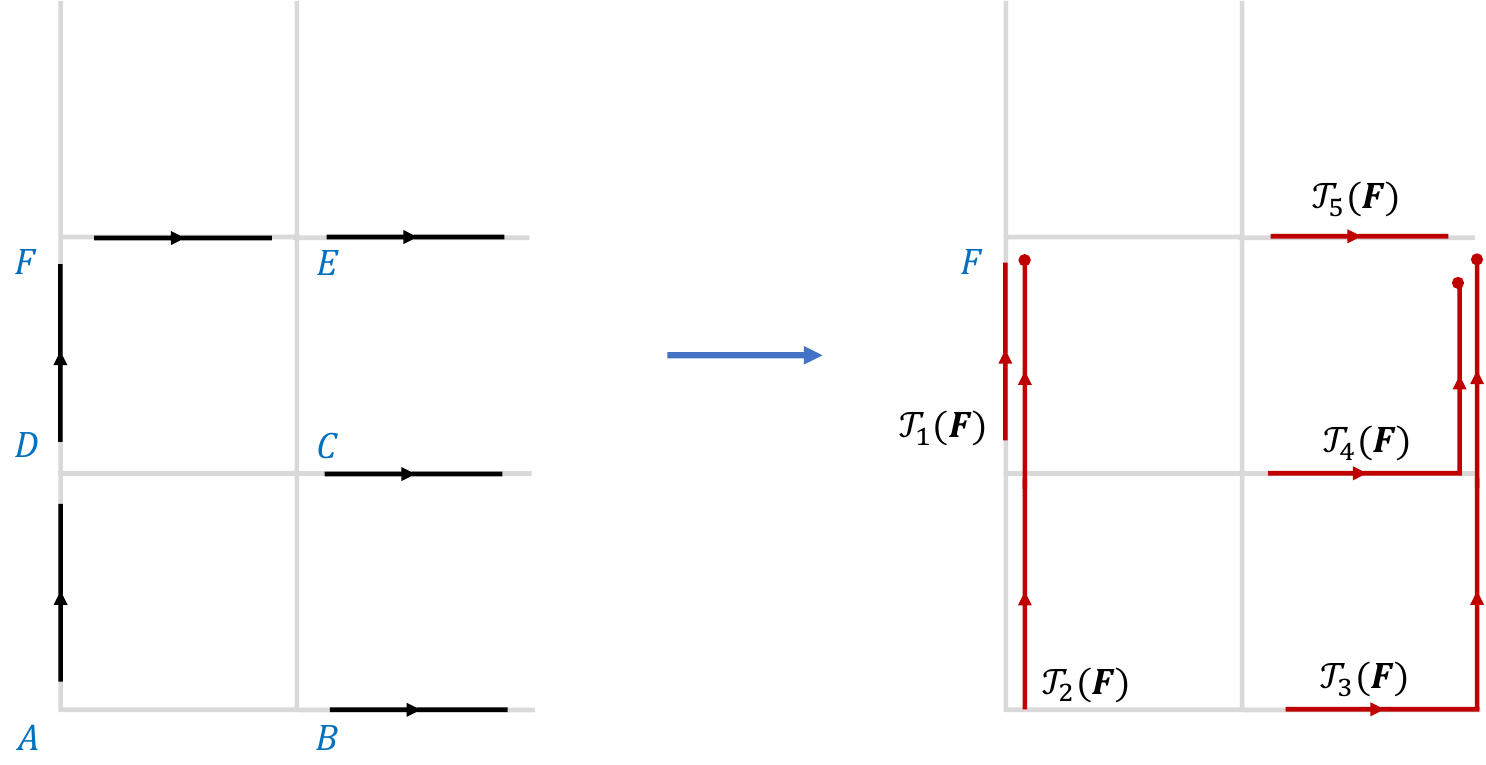}
    \caption{Graphical representation of canonical transformations to obtain five open strings $\mathcal{T}_i(\bm{F})$. In the final lattice, we report only the reference site $\bm{F}$.}
    \label{vertical_extension_CTs_2}
\end{figure}

We proceed by applying consecutive CTs, starting from the bottom left site $\bm{A}$ and following the lattice sites as for the single periodic plaquette. Graphically, the three sets of CTs are reported in Figs. \ref{vertical_extension_CTs_1}, \ref{vertical_extension_CTs_2}. As a counting of operations, we have 3 CTs for each plaquette, 2 CTs for $\mathcal{L}_{\hat{y}}(\bm{A})$, a single one for $\mathcal{L}_{\hat{x}}(\bm{E})$ and 4 CTs to get the open strings $\mathcal{T}_{i}(\bm{F})$, for a total of 22 CTs. 

By looking at Fig. \ref{vertical_extension_CTs_1}, we adopt the usual links order for the leftmost plaquette column, while for all the others we choose the origin in correspondence of the top left site of the square. For the reference site $\bm{F}$, we do not have an independent plaquette due to the PBCs. Regarding the large loops, they are centered in $\bm{A}$ (bottom left corner) and $\bm{E}$ (top right corner), with the same orientation as in the single plaquette case. Finally, in Fig. \ref{vertical_extension_CTs_2} we show how to transform all the remaining links into open strings ending in the reference point $\bm{F}$, which is our dependent point, i.e., the Gauss law $\mathcal{G}(\bm{F})$ is written in terms of the independent ones. The number of strings is equal to the number of independent Gauss laws, and we can use this freedom to fix them to the identity in the gauge group. As commented in the main text, this corresponds to the choice of a particular maximal tree gauge associated to our prescription \cite{Creutz1977,Ligterink2000}.

\subsection{\label{horizontal_plaquette}Horizontal extension}
\begin{figure}[h]
    \centering
    \includegraphics[width=0.45\linewidth]{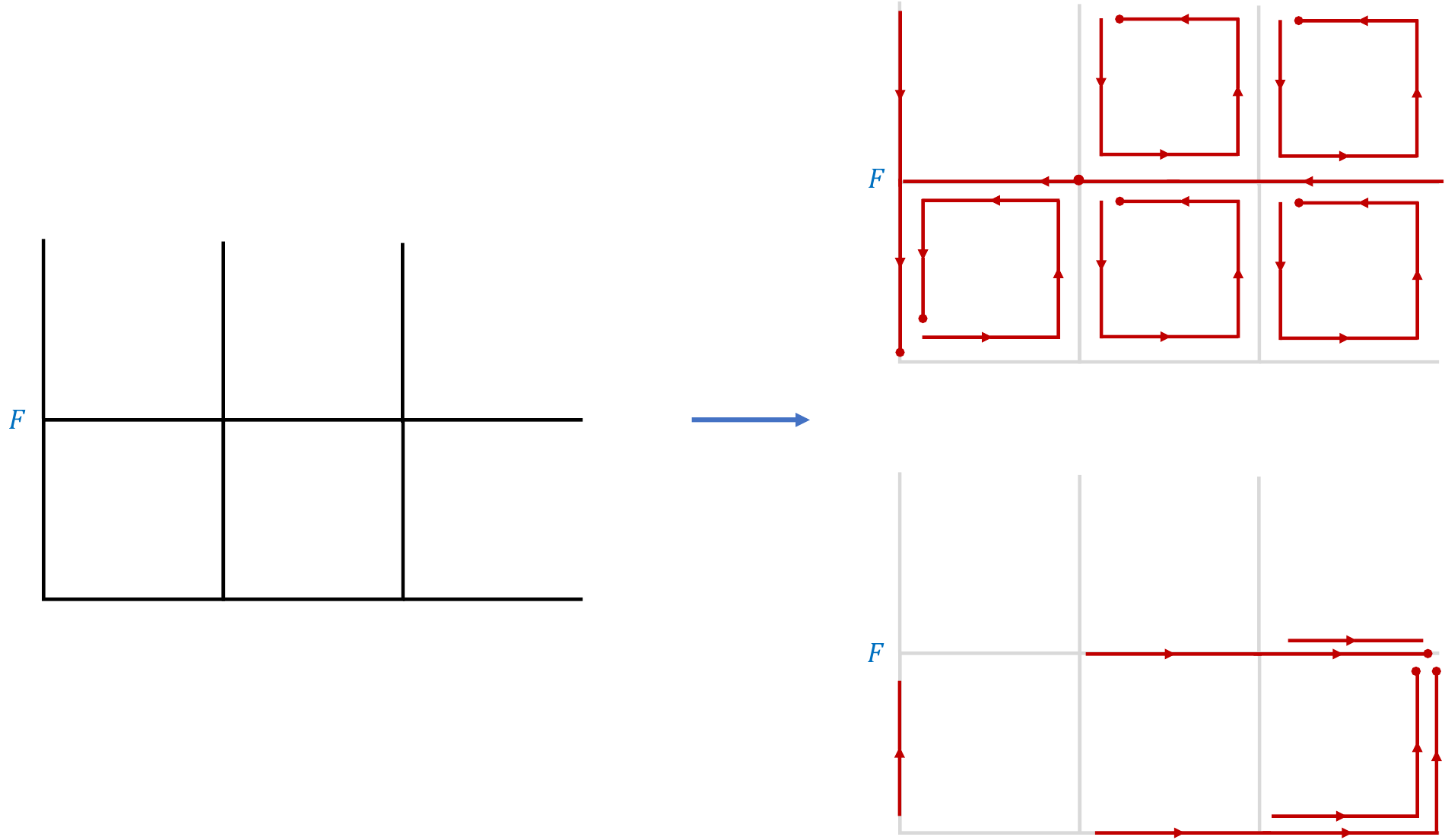}
    \caption{Graphical representation of the change of variables for the horizontal extension.  In the top lattice on the right, we show in red the plaquette and large loop closed strings, while in the bottom lattice we show, always in red, the open strings ending in the reference point. In all the lattices, we report only the reference site $\bm{F}$.}
    \label{horizontal_extension_CTs}
\end{figure}
The same counting of degrees of freedom done for the vertical extension applies here, and we just summarize the sets of CTs in Fig. \ref{horizontal_extension_CTs}. The strategy is the same: we apply iteratively single CTs to get independent plaquettes and large loops, and then build open strings ending in the reference point $\bm{F}$ in Fig. \ref{horizontal_extension_CTs}. 

\subsection{Full extension}
\begin{figure}[h]
    \centering
    \includegraphics[width=0.55\linewidth]{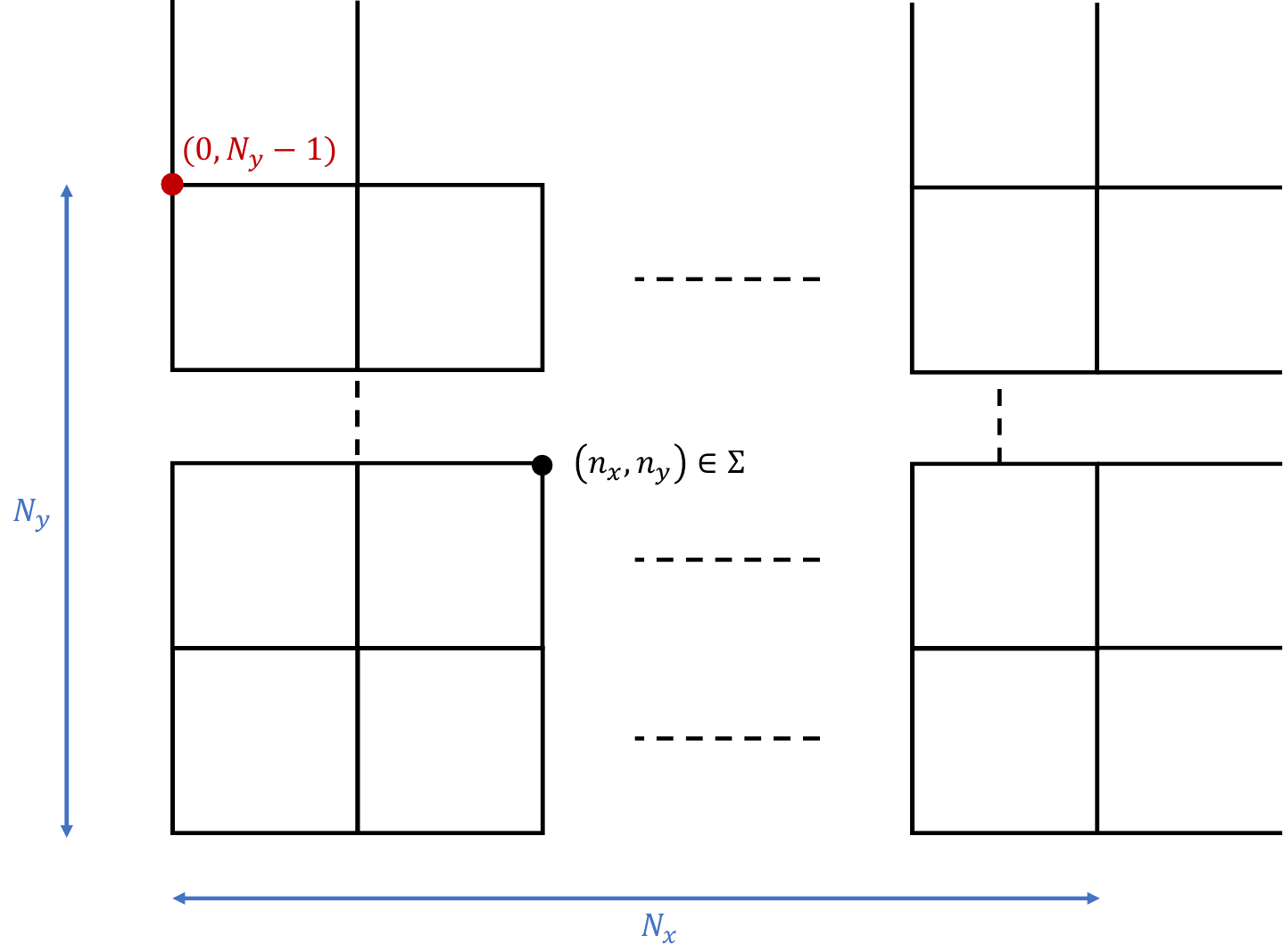}
    \caption{Unfolded torus $\Sigma$ in the two-dimensional plane, of size $N_x\times N_y$. A generic lattice site $(n_x,n_y)\in\Sigma$ if $n_{x,y}=0,\ldots,N_{x,y}-1$. In red we report the reference site $\bm{N}=(0,N_y-1)$ using this enumeration.}
    \label{full_ext_notation}
\end{figure}
Given the extension of Subsecs. \ref{vertical_plaquettes} and \ref{horizontal_plaquette}, we generalize the single periodic plaquette to an arbitrary $N_x\times N_y$ torus $\Sigma$. We set the notation for the lattice sites graphically in Fig. \ref{full_ext_notation}. In the standard formulation, we have $2N_xN_y$ links and $N_xN_y-1$ independent Gauss laws, for a total of $N_xN_y+1$ independent degrees of freedom. Once performed the full set of CTs, we end up in $N_xN_y-1$ plaquettes, i.e. a plaquette per lattice site except for the reference site $\bm{N}=(0,N_y-1)$, and 2 large loops wrapping the lattice, for a total of $N_xN_y+1$ physical independent degrees of freedom. The number of open strings can be obtained from the property that the single CT preserves the number of initial links. Therefore, from the counting of the loops, the number of open strings is $2N_xN_y-(N_xN_y+1)=N_xN_y-1$, i.e., exactly the number of independent Gauss law constraints.

In summary, if we call with $\mathcal{W}$ the set of closed loops and $\mathcal{S}$ the set of open strings, after the CTs we end up in
\begin{equation}
    \{U_{\mu}(\bm{n})\}_{\bm{n}\in\Sigma}\qquad\rightarrow\qquad\mathcal{W}\oplus\mathcal{S},\qquad |\mathcal{W}|=N_xN_y+1,\qquad|\mathcal{S}|=N_xN_y-1.
    \label{final_DOFs_counting}
\end{equation}
Only the loop electric fields are the physical ones, i.e., those that appear in the Hamiltonian and rule the dynamics of the theory after the explicit resolution of the Gauss laws, in complete analogy with the single periodic plaquette. Also in this case the number of CTs can be computed depending on the torus sizes \footnote{For every plaquette we have 3 CTs, and for the loops we have
\begin{equation}
    (N_x-1)\;\text{CTs for the}\;\hat{x}-\text{loop},\qquad(N_y-1)\;\text{CTs for the}\;\hat{y}-\text{loop},
\end{equation}
while for the open strings we have to distinguish
\begin{equation}
    \begin{cases}
        \text{open horizontal/vertical strings}:\qquad(N_{x,y}-2)\;\text{CTs to make a string of length} (N_{x,y}-1)\\
        \text{other open strings}:\qquad\sum_{j_y=1}^{Ny-1}\sum_{j_x=1}^{N_x-1}(N_x-j_x)(N_y-j_y)\;\text{total CTs}
    \end{cases}
\end{equation}
}.
\begin{figure}[h]
    \centering
    \includegraphics[width=0.4\linewidth]{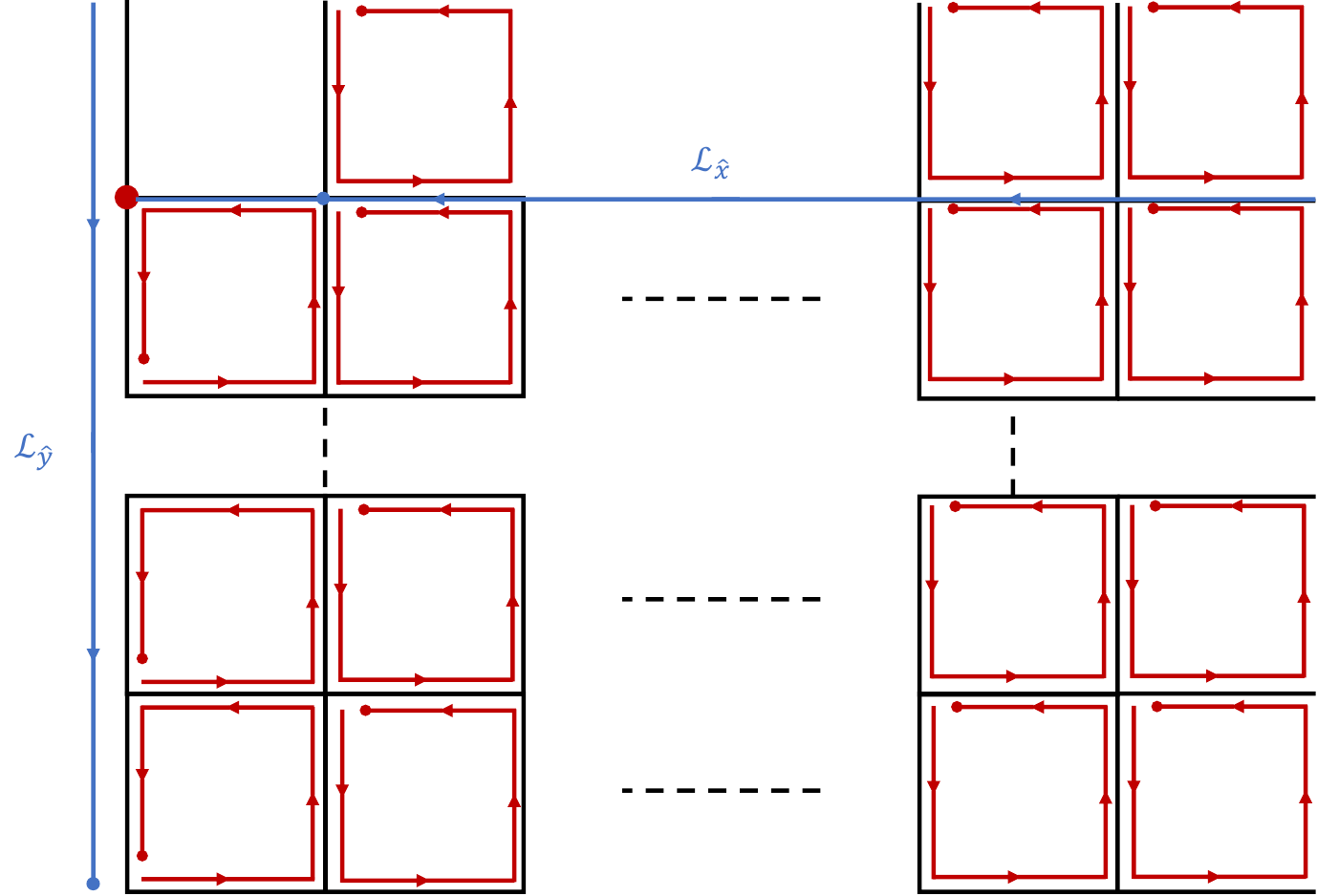}
    \hspace{3cm}
    \includegraphics[width=0.4\linewidth]{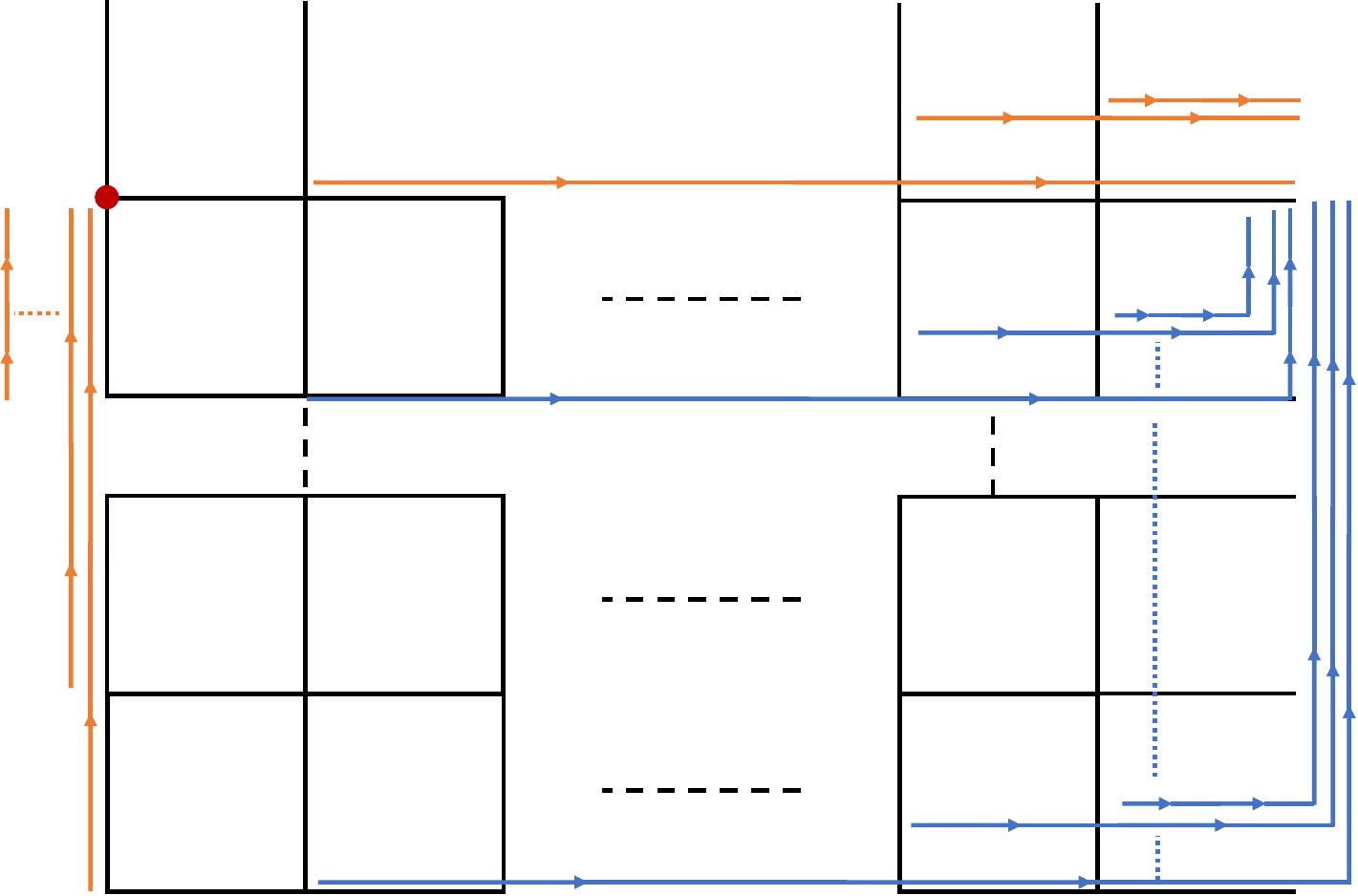}
    \caption{Graphical representation of reformulated degrees of freedom, the set $\mathcal{W}$ and $\mathcal{S}$. Left plot: we show the plaquettes (red strings) and the large loops (blue strings), and we refer to the dots as their origin. We observe that the leftmost column has standard plaquette definition, while for all the others we have to change origin to consistently complete the CTs. Right plot: vertical and horizontal open strings (in orange), and low-right corner-like strings (in blue).}
    \label{fullext_newDOFs}
\end{figure}

The graphical reformulated degrees of freedom are reported in Fig. \ref{fullext_newDOFs}. Besides the physical set $\mathcal{W}$, we observe that the most intricate part resides in the set of open strings $\mathcal{S}$. All of them end in the reference point $\bm{N}$, and their number is $N_x+N_y-2$ (orange strings in Fig. \ref{fullext_newDOFs}) plus $N_xN_y-N_x-N_y+1$ (blue strings in Fig. \ref{fullext_newDOFs}). The expected total is $|\mathcal{S}|=N_xN_y-1$, coincident with the number of independent Gauss laws. We set $\mathcal{T}_i\in\mathcal{S}$ to the group identity using the gauge freedom, analogously to the single periodic plaquette case.

%%%%%%%%%%%%%%%%%%%%%%%%%%%%%%%%%%%%%%%%%%%%%%%%%%%%%%%%%%%%%%%%%%%%%%%%%%%%

\section{\label{SU2_grouptheory_recap}Axis-angle coordinates and Haar measure of $SU(2)$}
We give here a brief reminder on $SU(2)$ axis-angle coordinates and representations of the operators needed in the main text. In this coordinate frame, group elements $|g\rangle\in SU(2)$ are interpreted as 3D rotations of angle $\omega$ and around a given axis $\hat{n}$, acting on the group generators \cite{Khersonskii1988,Bauer2023}. In spherical coordinates ${\bf \Omega}=(\omega,\theta,\phi)$, we have
\begin{equation}
    \hat{n}=\begin{pmatrix}
        \cos\phi\sin\theta\\
        \sin\phi\sin\theta\\
        \cos\theta
    \end{pmatrix},
    \qquad \theta\in[0,\pi],\qquad\phi\in[0,2\pi],\qquad \omega\in[0,2\pi].
\end{equation}
The fundamental, or defining, representation $j=1/2$ is a $2\times 2$ matrix, whose explicit expression is
\begin{equation}
    D^{\frac{1}{2}}({\bf \Omega})\equiv D({\bf \Omega})=
    \begin{pmatrix}
        \cos\frac{\omega}{2}-i\sin\frac{\omega}{2}\cos\theta && -i\sin\frac{\omega}{2}\sin\theta e^{-i\phi}\\\\
        -i\sin\frac{\omega}{2}\sin\theta e^{i\phi} && \cos\frac{\omega}{2}+i\sin\frac{\omega}{2}\cos\theta
    \end{pmatrix}.
    \label{axis_angle_defining_rep}
\end{equation}
We also remind that the Haar measure in these coordinates is 
\begin{equation}
    d\mu({\bf \Omega})=4\sin^2\frac{\omega}{2}\sin\theta \, d\omega\,d\theta\,d\phi,
    \label{axis_angle_haar_measure}
\end{equation}
and the volume of the group is $|\mathcal{G}|=16\pi^2$.

Concerning the representation of left and right electric operators, they are differential operators defined as
\begin{equation}
    {\bf \mathcal{E}}_{L,R}({\bf \Omega})=-{\bf \Sigma}\pm{\bf L},
    \label{electric_field_axisangle}
\end{equation}
where ${\bf L}$ is the angular momentum differential operator
\begin{equation}
    L_x=i\bigg(\sin\phi\frac{\partial}{\partial\theta}+\cot\theta\cos\phi\frac{\partial}{\partial\phi}\bigg),\qquad L_y=i\bigg(-\cos\phi\frac{\partial}{\partial\theta}+\cot\theta\sin\phi\frac{\partial}{\partial\phi}\bigg),\qquad L_z=-i\frac{\partial}{\partial\phi}
\end{equation}
and ${\bf \Sigma}$ has components
\begin{equation}
    \Sigma_x=2i\sin\theta\cos\phi\frac{\partial}{\partial\omega}+i\cot\frac{\omega}{2}\bigg(\cos\theta\cos\phi\frac{\partial}{\partial\theta}-\csc\theta\sin\phi\frac{\partial}{\partial\phi}\bigg),
\end{equation}
\begin{equation}
    \Sigma_y=2i\sin\theta\sin\phi\frac{\partial}{\partial\omega}+i\cot\frac{\omega}{2}\bigg(\cos\theta\sin\phi\frac{\partial}{\partial\theta}+\csc\theta\cos\phi\frac{\partial}{\partial\phi}\bigg),
\end{equation}
\begin{equation}
    \Sigma_z= 2i\cos\theta\frac{\partial}{\partial\omega}-i\cot\frac{\omega}{2}\sin\theta\frac{\partial}{\partial\theta}.
\end{equation}
Consequently, we have the square electric operator 
\begin{equation}
    \mathcal{E}^2({\bf \Omega})=\frac{{\bf \hat{L}}^2}{\sin^2\frac{\omega}{2}}-4\bigg[\frac{\partial^2}{\partial\omega^2}+\cot\frac{\omega}{2}\frac{\partial}{\partial\omega}\bigg].
    \label{squareE_operator_axisangle}
\end{equation}

Regarding the loop operators in the group basis, they can be represented using the unitary matrices $D^j_{mm'}(g)$ associated to a specific irreducible representation (irrep) of $SU(2)$ \cite{serre1977linear,ZoharPRD2015}. Therefore, in general, the loop operator in the $j$-irrep is
\begin{equation}
    W^j_{mm'}(i)=\int\;d\mu(g_i)\;D^j_{mm'}(g_i)|g_i\rangle\langle g_i|,
    \label{loop_operator_group_basis}
\end{equation}
being $i$ the loop index of $\mathcal{L}_i\in\mathcal{W}$. The loop operators are diagonal in the group basis. In the whole manuscript we always fix $j=1/2$ and consider the loop operators in the defining representation.

%%%%%%%%%%%%%%%%%%%%%%%%%%%%%%%%%%%%%%%%%%%%%%%%%%%%%%%%%%%%%%%%%%%%%%%%%%%%

\section{\label{WE_electric_field_computations}Left loop electric operator reduced matrix elements}
We compute the reduced matrix element of the left loop electric field by decomposing it into the spherical vectors of ${\bf {L}}$ and ${\bf {\Sigma}}$. For the first one we fix $m_I=1,\;m_J=0$ and compute it for the $q=1$ component, i.e. ${\bf {L}^+}$, as
\begin{equation}
    \langle\alpha_{\ell_I}\ell_I||{\bf {L}}||\alpha_{\ell_J}\ell_J\rangle=\frac{\langle\alpha_{\ell_I},\ell_I,1|{\bf {L}}^+|\alpha_{\ell_J},\ell_J,0\rangle}{\braket{\ell_I1|\ell_J 0,11}}.
\end{equation}
Explicitly this is
\begin{equation}
    \langle\alpha_{\ell_I},\ell_I,1|{\bf {L}}|\alpha_{\ell_J},\ell_J,0\rangle=\int\;d\omega\;u^*_{\alpha_{\ell_I}}(\omega)u_{\alpha_{\ell_J}}(\omega)Y^*_{\ell_I1}(\theta,\phi){\bf {L}}^+Y_{\ell_J0}(\theta,\phi)\;\sin\theta d\theta d\phi,
\end{equation}
and by taking into account that ${\bf {L}}^+Y_{\ell_J0}(\theta,\phi)=\sqrt{\ell_J(\ell_J+1)}Y_{\ell_J1}(\theta,\phi)$ we get $\langle\alpha_{\ell_I},\ell_I,1|{\bf {L}}^+|\alpha_{\ell_J},\ell_J,0\rangle=\delta_{\ell_I,\ell_J}\sqrt{\ell_J(\ell_J+1)}$. Considering also that $\braket{\ell_I1|\ell_J 0,11}=-1/\sqrt{2}$, the reduced matrix element is
\begin{equation}
    \langle\alpha_{\ell_I}\ell_I||{\bf {L}}||\alpha_{\ell_J}\ell_J\rangle=-\sqrt{2\ell_J(\ell_J+1)}\delta_{\ell_I,\ell_J}.
    \label{reduced_element_L}
\end{equation}
We proceed equally for the spherical vector of ${\bf {\Sigma}}$, and here we fix $m_I=m_J=0$ and $q=0$, computing the reduced matrix element as
\begin{equation}
    \langle\alpha_{\ell_I}\ell_I||{\bf {\Sigma}}||\alpha_{\ell_J}\ell_J\rangle=\frac{\langle\alpha_{\ell_I},\ell_I,0|{\bf {\Sigma}}^0|\alpha_{\ell_J},\ell_J,0\rangle}{\braket{\ell_I0|\ell_J 0,10}}.
\end{equation}
In the integral representation, the numerator on the right-hand side is
\begin{align}
    \nonumber
    \langle\alpha_{\ell_I},\ell_I,0|{\bf {\Sigma}}^0|\alpha_{\ell_J},\ell_J,0\rangle=&i\int\;d\omega\;u^*_{\alpha_{\ell_I}}(\omega)Y^*_{\ell_I0}(\theta,\phi)\bigg[2\cos\theta\bigg(\frac{\partial}{\partial\omega}-\frac{1}{2}\cot\frac{\omega}{2}\bigg)-\cot\frac{\omega}{2}\sin\theta\frac{\partial}{\partial\theta}\bigg]\\
    &\times Y_{\ell_J0}(\theta,\phi)u_{\alpha_{\ell_J}}(\omega)\;\sin\theta d\theta d\phi,
\end{align}
and we have the properties of spherical harmonics \cite{Khersonskii1988}
\begin{equation}
    \cos\theta\;Y_{\ell_J 0}(\theta,\phi)=\frac{\ell_J+1}{\sqrt{(2\ell_J+1)(2\ell_J+3)}}Y_{\ell_J+1,0}(\theta,\phi)+\frac{\ell_J}{\sqrt{(2\ell_J-1)(2\ell_J+1)}}Y_{\ell_J-1,0}(\theta,\phi),
    \label{cos_Y_product}
\end{equation}
\begin{equation}
    \sin\theta\;\frac{\partial Y_{\ell_J 0}(\theta,\phi)}{\partial\theta}=\ell_J(\ell_J+1)\bigg[\frac{Y_{\ell_J+1,0}(\theta,\phi)}{\sqrt{(2\ell_J+1)(2\ell_J+3)}}-\frac{Y_{\ell_J-1,0}(\theta,\phi)}{\sqrt{(2\ell_J+1)(2\ell_J-1)}}\bigg].
\end{equation}
Using the normalization of the spherical harmonics with respect to the angular measure $d\Omega\equiv\sin\theta d\theta d\phi$, we end up in
\begin{align}
    \nonumber
    \langle\alpha_{\ell_I},\ell_I,0|{\bf {\Sigma}}^0|\alpha_{\ell_J},\ell_J,0\rangle=&i\int\;d\omega\;u^*_{\alpha_{\ell_I}}(\omega)\bigg[\frac{2(\ell_J+1)}{\sqrt{(2\ell_J+1)(2\ell_J+3)}}\bigg(\frac{\partial}{\partial\omega}-\frac{\ell_J+1}{2}\cot\frac{\omega}{2}\bigg)\delta_{\ell_J+1,\ell_I}\\
    &+\frac{2\ell_J}{\sqrt{(2\ell_J+1)(2\ell_J-1)}}\bigg(\frac{\partial}{\partial\omega}+\frac{\ell_J}{2}\cot\frac{\omega}{2}\bigg)\delta_{\ell_J-1,\ell_I}\bigg]u_{\alpha_{\ell_J}}(\omega).
\end{align}
The involved Clebsch--Gordan coefficients 
\begin{equation}
    \braket{\ell_J+1,0|\ell_J 0,10}=\frac{\ell_J+1}{\sqrt{(\ell_J+1)(2\ell_J+1)}},\qquad  \braket{\ell_J-1,0|\ell_J 0,10}=-\frac{\ell_J}{\sqrt{\ell_J(2\ell_J+1)}},
\end{equation}
lead to the simplification
\begin{equation}
    \frac{2(\ell_J+1)}{\sqrt{(2\ell_J+1)(2\ell_J+3)}}\cdot\frac{1}{\braket{\ell_J+1,0|\ell_J 0,10}}=2\sqrt{\frac{\ell_J+1}{2(\ell_J+1)+1}},
    \label{CG_simplification_l-1_l}
\end{equation}
\begin{equation}
    \frac{2\ell_J}{\sqrt{(2\ell_J+1)(2\ell_J-1)}}\cdot\frac{1}{\braket{\ell_J-1,0|\ell_J 0,10}}=-2\sqrt{\frac{\ell_J}{2\ell_J-1}}.
    \label{CG_simplification_l+1_l}
\end{equation}
Finally, the reduced matrix element can be written as
\begin{align}
    \nonumber
    \langle\alpha_{\ell_I}\ell_I||{\bf {\Sigma}}||\alpha_{\ell_J}\ell_J\rangle=&2i\int\;d\omega\;u^*_{\alpha_{\ell_I}}(\omega)\bigg[\sqrt{\frac{\ell_J+1}{2(\ell_J+1)+1}}\bigg(\frac{\partial}{\partial\omega}-\frac{\ell_J+1}{2}\cot\frac{\omega}{2}\bigg)\delta_{\ell_J+1,\ell_I}\\
    &-\sqrt{\frac{\ell_J}{2\ell_J-1}}\bigg(\frac{\partial}{\partial\omega}+\frac{\ell_J}{2}\cot\frac{\omega}{2}\bigg)\delta_{\ell_J-1,\ell_I}\bigg]u_{\alpha_{\ell_J}}(\omega).
    \label{reduced_element_Sigma}
\end{align}
The reduced element of the left loop electric field is the sum of the two Eqs. \eqref{reduced_element_L}, \eqref{reduced_element_Sigma} with the relative signs dictated by the definition of Eq. \eqref{electric_field_axisangle}.

%%%%%%%%%%%%%%%%%%%%%%%%%%%%%%%%%%%%%%%%%%%%%%%%%%%%%%%%%%%%%%%%%%%%%%%%%%%%

\section{\label{WE_EL_ER_product_computations}Left-right loop electric operators product matrix elements}
In the dual Hamiltonian we need to evaluate the scalar term $\mathcal{E}^a_L(\bm{B})\mathcal{E}^a_R(\bm{B})$. Since this is the only term with product of left and right electric fields, we omit the site index $\bm{B}$ to lighten the notation. Starting from the definition of left and right electric fields in Eq. \eqref{electric_field_axisangle}, and given that $[{\bf L},{\bf \Sigma}]=0$, we have
\begin{equation}
    \mathcal{E}^a_L\mathcal{E}^a_R={\bf \Sigma}^2-{\bf L}^2.
\end{equation}
As ${\bf L}^2 Y_{\ell m}(\theta,\phi)=\ell(\ell+1)Y_{\ell m}(\theta,\phi)$ by definition, we need to evaluate the action of ${\bf \Sigma}^2$ on a given state $|\alpha_{\ell},\ell,m\rangle$. In axis-angle coordinates, we can write 
\begin{equation}
    {\bf \Sigma}^2=-\cot^2\frac{\omega}{2}\bigg(\cot\theta\frac{\partial}{\partial\theta}+\frac{\partial^2}{\partial\theta^2}+\csc^2\theta\frac{\partial}{\partial\phi^2}\bigg)-\frac{2}{\sin^2\frac{\omega}{2}}\bigg(\sin\omega\frac{\partial}{\partial\omega}-(\cos\omega-1)\frac{\partial^2}{\partial\omega^2}\bigg)\equiv{\bf \Sigma}_A^2({\bf \Omega})+{\bf \Sigma}_B^2(\omega)
\end{equation}
and evaluate each contribution separately. For ${\bf \Sigma}_A^2$ we recognize that 
\begin{equation}
    \cot\theta\frac{\partial}{\partial\theta}+\frac{\partial^2}{\partial\theta^2}+\csc^2\theta\frac{\partial}{\partial\phi^2}=\frac{1}{\sin\theta}\frac{\partial}{\partial\theta}\bigg(\sin\theta\frac{\partial}{\partial\theta}\bigg)+\csc^2\theta\frac{\partial}{\partial\phi^2}\equiv-{\bf L}^2(\theta,\phi),
\end{equation}
therefore we get immediately
\begin{equation}
    \langle\alpha_{\ell_I},\ell_I,m_I|{\bf \Sigma}_A^2|\alpha_{\ell_J},\ell_J,m_J\rangle=\delta_{\ell_I,\ell_J}\delta_{m_I,m_J}\ell_J(\ell_J+1)\int\;d\omega\;u^*_{\alpha_{\ell_I}}(\omega)\cot^2\frac{\omega}{2}u_{\alpha_{\ell_J}}(\omega).
\end{equation}
Regarding ${\bf \Sigma}_B^2$, since it depends only on the axis coordinate $\omega$, we can directly evaluate the integral, obtaining  
\begin{equation}
    \langle\alpha_{\ell_I},\ell_I,m_I|{\bf \Sigma}_B^2|\alpha_{\ell_J},\ell_J,m_J\rangle=-\delta_{\ell_I,\ell_J}\delta_{m_I,m_J}\int\;d\omega\;u^*_{\alpha_{\ell_I}}(\omega)\bigg(1+4\frac{\partial^2}{\partial\omega^2}\bigg)u_{\alpha_{\ell_J}}(\omega).
\end{equation}
We consequently get
\begin{align}
    \nonumber
    \langle\alpha_{\ell_I},\ell_I,m_I|\mathcal{E}^a_L\mathcal{E}^a_R|\alpha_{\ell_J},\ell_J,m_J\rangle&=-\delta_{\ell_I,\ell_J}\delta_{m_I,m_J}\bigg[\ell_J(\ell_J+1)\\
    &+\int\;d\omega\;u^*_{\alpha_{\ell_I}}(\omega)\bigg(1+4\frac{\partial^2}{\partial\omega^2}-\ell_J(\ell_J+1)\cot^2\frac{\omega}{2}\bigg)u_{\alpha_{\ell_J}}(\omega)\bigg].
    \label{left-right_product_Efields_WE}
\end{align}

%%%%%%%%%%%%%%%%%%%%%%%%%%%%%%%%%%%%%%%%%%%%%%%%%%%%%%%%%%%%%%%%%%%%%%%%%%%%

\section{\label{WE_loop_computations}Loop operator reduced matrix elements}
We report here the computation of the loop operator reduced matrix elements, for the application of the Wigner--Eckart theorem. To this end, we decompose the loop operator as \cite{Bauer2023}
\begin{equation}
    W\equiv S\mathbb{1}+W^aT^a,\qquad S\equiv\frac{\text{Tr}W}{2},\qquad W^a\equiv2\text{Tr}[T^aW],
    \label{spherical_decomposition_loopW}
\end{equation}
where $T^a=\sigma^a/2$. The scalar part $S$ is straightforward to compute from Eq. \eqref{axis_angle_defining_rep}, and equals to
\begin{equation}
    \langle\alpha_{\ell_I},\ell_I,m_I|S|\alpha_{\ell_J},\ell_J,m_J\rangle=\delta_{\ell_I,\ell_J}\delta_{m_I,m_J}\int\;d\omega\;u^*_{\alpha_{\ell_I}}(\omega)\cos\frac{\omega}{2}u_{\alpha_{\ell_J}}(\omega).
    \label{scalar_loopW_WEtheorem}
\end{equation}
Concerning the vector part, we focus on the $q=0$ spherical component of $W^a$, i.e.,
\begin{equation}
    2\text{Tr}(T^z W)=\text{Tr}(\sigma^z W)=-2i\cos\theta\sin\frac{\omega}{2}
\end{equation}
and use the independence of the reduced matrix elements on the azimuthal quantum number to fix $m_I=m_J=0$. We then evaluate
\begin{equation}
    \langle\alpha_{\ell_I}\ell_I||2\text{Tr}({\bf T}W)||\alpha_{\ell_J}\ell_J\rangle=\frac{\langle\alpha_{\ell_I},\ell_I,0|\text{Tr}(\sigma^z W)|\alpha_{\ell_J},\ell_J,0\rangle}{\braket{\ell_I0|\ell_J 0,10}}.
\end{equation}
The numerator on the right-hand side can be written immediately by writing explicitly the value $\cos\theta\;Y_{\ell_J,0}(\theta,\phi)$, using Eq. \eqref{cos_Y_product}, as
\begin{align}
    \nonumber
    \langle\alpha_{\ell_I},\ell_I,0|\text{Tr}(\sigma^zW)|\alpha_{\ell_J},\ell_J,0\rangle=&-2i\int\;d\omega\;u_{\alpha_{\ell_I}}^*(\omega)\sin\frac{\omega}{2}u_{\alpha_{\ell_J}}(\omega)\bigg[\frac{\ell_J+1}{\sqrt{(2\ell_J+1)(2\ell_J+3)}}\delta_{\ell_J+1,\ell_I}\\
    &+\frac{\ell_J}{\sqrt{(2\ell_J+1)(2\ell_J-1)}}\delta_{\ell_J-1,\ell_I}\bigg].
\end{align}
The reduced matrix element is obtained as the ratio of this last expression to the corresponding Clebsch--Gordan coefficients. By simplifying as in Eqs. \eqref{CG_simplification_l-1_l}, \eqref{CG_simplification_l+1_l}, we get
\begin{align}
    \nonumber
    \langle\alpha_{\ell_I}\ell_I||2\text{Tr}({\bf T}W)||\alpha_{\ell_J}\ell_J\rangle=&-2i\int\;d\omega\;u_{\alpha_{\ell_I}}^*(\omega)\sin\frac{\omega}{2}u_{\alpha_{\ell_J}}(\omega)\bigg[\sqrt{\frac{\ell_J+1}{2(\ell_J+1)+1}}\delta_{\ell_J+1,\ell_I}\\
    &-\sqrt{\frac{\ell_J}{2\ell_J-1}}\delta_{\ell_J-1,\ell_I}\bigg].
    \label{vector_loopW_WEtheorem}
\end{align}
Finally, the total matrix element is given by summing Eqs. \eqref{scalar_loopW_WEtheorem} and \eqref{vector_loopW_WEtheorem}, weighted with the proper Clebsch--Gordan coefficient in light of the Wigner--Eckart theorem.

%%%%%%%%%%%%%%%%%%%%%%%%%%%%%%%%%%%%%%%%%%%%%%%%%%%%%%%%%%%%%%%%%%%%%%%%%%%%

\section{\label{Rab_scalar_vector_decomposition}Parallel transport decomposition}
We show how to write the parallel transport $\mathcal{R}^{ba}(W)=2\text{Tr}[W^\dagger T^b W T^a]$ of a generic loop $W$ in terms of scalar and vector components. If we decompose the loop operator $W$ as in Eq. \eqref{spherical_decomposition_loopW} we have, in the specific case of $SU(2)$, that
\begin{equation}
    \mathcal{R}^{ba}(W)=\frac{1}{2}\text{Tr}\bigg[S^*S\sigma^b\sigma^a+\frac{1}{2}\bigg(S^*W^d\sigma^b\sigma^d\sigma^a+S(W^*)^c\sigma^c\sigma^b\sigma^a\bigg)+\frac{1}{4}(W^*)^cW^d\sigma^c\sigma^b\sigma^d\sigma^a\bigg].
\end{equation}
Due to the properties of the Pauli matrices
\begin{equation}
    \text{Tr}[\sigma^b\sigma^a]=2\delta^{ba},\qquad\text{Tr}[\sigma^b\sigma^d\sigma^a]=2i\epsilon^{bda},\qquad\text{Tr}[\sigma^c\sigma^b\sigma^d\sigma^a]=2(\delta^{cb}\delta^{da}-\delta^{cd}\delta^{ba}+\delta^{ca}\delta^{bd}),
\end{equation}
we obtain
\begin{equation}
    \mathcal{R}^{ba}(W)=S^* S\delta^{ba}+\frac{i}{2}\epsilon^{cba}[W^* S-S^* W]^c+\frac{1}{4}[(W^*)^bW^a-(W^*)^c W^c\delta^{ba}+(W^*)^aW^b],
\end{equation}
where repeated indices are summed over. By noticing that $S=S^*=\cos(\omega/2)$, $(W^*-W)^c=-2\text{Im}(W^c)$ and $(W^*)^c W^c=4\sin^2(\omega/2)$, we can rewrite everything more compactly as
\begin{equation}
    \mathcal{R}^{ba}(W)=\cos\omega\;\delta^{ba}+\frac{1}{4}[(W^*)^bW^a+(W^*)^aW^b]-i\cos\frac{\omega}{2}\text{Im}(W^c)\epsilon^{cba}.
    \label{parallel_transport_scalar_vector_decomposition}
\end{equation}
In axis-angle coordinates we can write explicitly the last two contributions as three-dimensional matrices
\begin{equation}
    \frac{1}{4}[(W^*)^bW^a+(W^*)^aW^b]=2\sin^2\frac{\omega}{2}
    \begin{pmatrix}
        \sin^2\theta\cos^2\phi && \sin^2\theta\cos\phi\sin\phi && \sin\theta\cos\theta\cos\phi\\
        \sin^2\theta\cos\phi\sin\phi && \sin^2\theta \sin^2\phi && \sin\theta\cos\theta\sin\phi\\
        \sin\theta\cos\theta\cos\phi && \sin\theta\cos\theta\sin\phi && \cos^2\theta
    \end{pmatrix}
    \equiv 2\sin^2\frac{\omega}{2} R^{ab}_S,
    \label{Rab_symm_matrix}
\end{equation}
\begin{equation}
    -i\cos\frac{\omega}{2}\text{Im}(W^c)\epsilon^{cba}=i\sin\omega
    \begin{pmatrix}
        0 && -\cos\theta && \sin\theta\sin\phi \\
        \cos\theta && 0 && -\sin\theta\cos\phi \\
        -\sin\theta\sin\phi && \sin\theta\cos\phi && 0
    \end{pmatrix}
    \equiv i\sin\omega R^{ab}_A
    \label{Rab_antisymm_matrix}
\end{equation}
In the computation of matrix elements of the parallel transports in Eq. \eqref{parallel_transport_scalar_vector_decomposition}, the following integral appears
\begin{equation}
    \langle \alpha_{\ell_I},\ell_I,m_I|\mathcal{R}^{ba}(W)|\alpha_{\ell_J},\ell_J,m_J\rangle \ni \int\;\;Y^*_{\ell_I m_I}f_{ab}(\theta,\phi)Y_{\ell_J,m_J}\;\sin\theta d\theta d\phi,
\end{equation}
where $f_{ab}(\theta,\phi)$ is a combination of the matrix elements written above, i.e., a linear combination of trigonometric functions. To compute this, we write $f_{ab}(\theta,\phi)$ using spherical harmonics, expressing the integrand as the product of three spherical harmonics, and then use the relation \cite{sakurai1994modern}
\begin{equation}
    Y_{\ell m}Y_{\ell_J m_J}=\sum_{L=|\ell-\ell_J|}^{\ell+\ell_J}\sum_{M=-L}^L\sqrt{\frac{(2\ell+1)(2\ell_J+1)}{4\pi(2L+1)}}\langle\ell,0,\ell_J,0|L,0\rangle\langle \ell,m,\ell_J,m_J|L,M\rangle\;Y_{LM}
    \label{products_Ylm}
\end{equation}
to decompose the product of two spherical harmonics. In this way, the integral reduces to the normalization condition on the spherical harmonics, giving rise to selection rules in the angular momentum quantum numbers. For completeness, we report the expressions of the matrices in Eqs. \eqref{Rab_symm_matrix}, \eqref{Rab_antisymm_matrix}, in terms of spherical harmonics
\begin{equation}
    R^{ab}_S=
    \begin{pmatrix}
        \sqrt{\frac{2\pi}{15}}(Y_{2,2}+Y_{2,-2})+\frac{1}{3}(1-2\sqrt{\frac{\pi}{5}}Y_{2,0}) && i\sqrt{\frac{2\pi}{15}}(Y_{2,-2}-Y_{2,2}) && \sqrt{\frac{2\pi}{15}}(Y_{2,-1}-Y_{2,1})\\
        i\sqrt{\frac{2\pi}{15}}(Y_{2,-2}-Y_{2,2}) && \frac{1}{3}(1-2\sqrt{\frac{\pi}{5}}Y_{2,0})-\sqrt{\frac{2\pi}{15}}(Y_{2,2}+Y_{2,-2}) && i\sqrt{\frac{2\pi}{15}}(Y_{2,1}+Y_{2,-1})\\
        \sqrt{\frac{2\pi}{15}}(Y_{2,-1}-Y_{2,1}) && i\sqrt{\frac{2\pi}{15}}(Y_{2,1}+Y_{2,-1}) && \frac{1}{3}(1+4\sqrt{\frac{\pi}{5}}Y_{2,0})
    \end{pmatrix},
\end{equation}
\begin{equation}
    R^{ab}_A=
    \begin{pmatrix}
        0 && -2\sqrt{\frac{\pi}{3}}Y_{1,0} && i\sqrt{\frac{2\pi}{3}}(Y_{1,1}+Y_{1,-1})\\
        2\sqrt{\frac{\pi}{3}}Y_{1,0} && 0 && \sqrt{\frac{2\pi}{3}}(Y_{1,-1}-Y_{1,1})\\
        -i\sqrt{\frac{2\pi}{3}}(Y_{1,1}+Y_{1,-1}) && -\sqrt{\frac{2\pi}{3}}(Y_{1,-1}-Y_{1,1}) && 0
    \end{pmatrix}.
\end{equation}

\begin{figure*}[t!]
    \centering
    \includegraphics[width=0.95\linewidth]{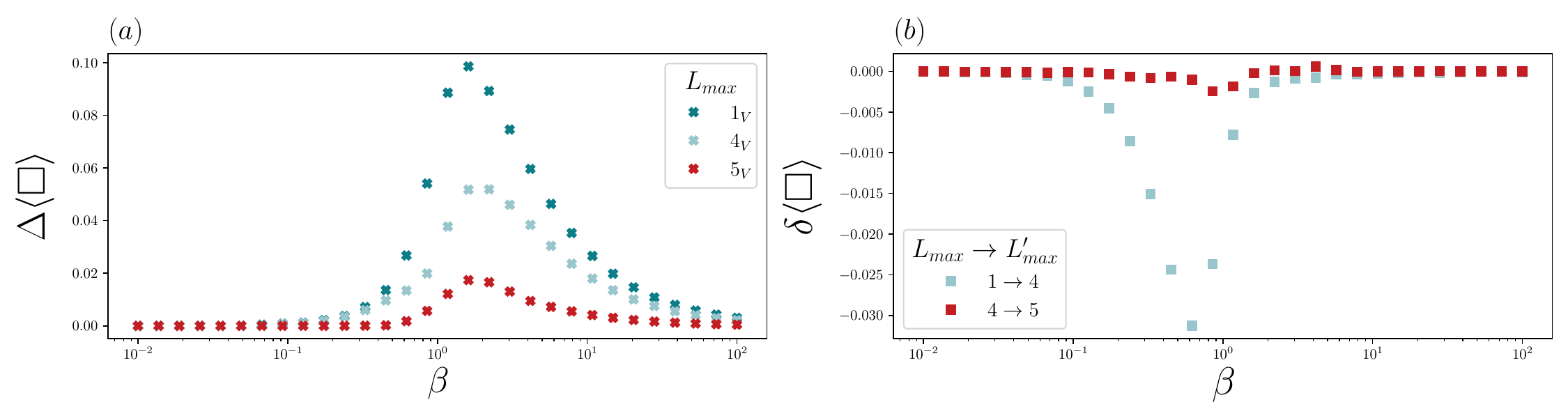}
    \caption{(a) Absolute difference $\Delta \langle\square\rangle\equiv\langle\square\rangle(\bm{g_0})-\langle\square\rangle(\bm{g_V})$ in the expectation values of the plaquette operator between the initial ansatz $\bm{g}_0$ and the optimal couplings as a function of the bare coupling $\beta$, for the different values of truncations $L_{max}=1,4,5$. (b) Absolute difference in the expectation value of the plaquette operator $\delta \langle\square\rangle\equiv\langle\square\rangle_{L_{max}'}-\langle\square\rangle_{L_{max}}$ between consecutive values of the truncation $L_{max}$ as a function of the bare coupling $\beta$.}
    \label{absolute_error_plaquette_operator}
\end{figure*}

\section{\label{relative_plaquette} Additional discussion on the plaquette operator}
In this Appendix we present additional data regarding the determination of the plaquette operator. Due to the renormalization of the coupling, in order to extract continuum-limit results from a LGT it is important to employ a truncation scheme that is efficient for all values of the coupling constants. Therefore, it has been our goal to assess the performance of the variationally optimized basis in \textit{all} the range of coupling strengths. For this reason, in Fig. \ref{running_couplings_plot}d) we have decided to show the relative difference in the determination of the plaquette operator, $\frac{\Delta \langle\square\rangle}{\langle\square\rangle}$. However, due to the bounded nature of $\langle\square\rangle$ itself, it is also interesting to observe the absolute error in the determination of the plaquette expectation value. We report it, for completeness, in Fig. \ref{absolute_error_plaquette_operator}. Since the plaquette expectation value vanishes for $\beta\rightarrow\infty$ the computation of $\frac{\Delta \langle\square\rangle}{\langle\square\rangle}$ is demanding truncation-wise and especially sensitive to the numerical precision of the computation. Since in Fig. \ref{running_couplings_plot}d) this quantity is shown to behave smoothly towards weak coupling, we can deduce that the truncation performs well in the determination of the plaquette $\langle\square\rangle$ and that the difference $\Delta \langle\square\rangle$ is not plagued by numerical errors in the computed range of the coupling strength.

\section{\label{numerical_implementation}Numerical implementation}
Numerical data has been obtained using exact diagonalization (ED). All the numerical work was done with Mathematica \cite{Mathematica}. The local eigenstates were obtained with a finite element method with a precision $\delta=10^{-2}$. All Hamiltonian matrix elements were obtained through numerical integration. The methods ``DoubleExponential" and ``SymbolicPreprocessing" were found to give the best performance for the types of integrals involved in this model. To find the optimal choice for the local basis, the minimization of the energy was performed using an interior point method \cite{InteriorPoint}. Convergence was assumed when the relative difference in the energy $\delta E/E\leq 10^{-4}$.

%%%%%%%%%%%%%%%%%%%%%%%%%%%%%%%%%%%%%%%%%%%%%%%%%%%%%%%%%%%%%%%%%%%%%%%%%%%%

\bibliography{biblio}

%%%%%%%%%%%%%%%%%%%%%%%%%%%%%%%%%%%%%%%%%%%%%%%%%%%%%%%%%%%%%%%%%%%%%%%%%%%%

\end{document}